\documentclass[twocolumn,tighten]{aastex62}
\pdfoutput=1 
\usepackage{amsmath,amstext}
\usepackage[T1]{fontenc}
\usepackage{ae,aecompl}
\usepackage[utf8]{inputenc}
\usepackage{apjfonts} 
\usepackage[figure,figure*]{hypcap}
\usepackage{enumitem}
\usepackage{bm}
\input{hyperlink-year-only-natbib-patch}

\newcommand*{\http}[1]{\href{http://#1}{#1}}
\newcommand*{\https}[1]{\href{https://#1}{#1}}

\newcommand{\msun}{\ensuremath{{M_{\odot}}}}

\shorttitle{Modeling the Subhalo--Satellite Connection}
\shortauthors{Nadler et al.}

\begin{document}

\title{Modeling the Connection Between Subhalos and Satellites in Milky Way-like Systems}

\author[0000-0002-1182-3825]{Ethan~O.~Nadler}
\affiliation{Kavli Institute for Particle Astrophysics and Cosmology and Department of Physics, Stanford University, Stanford, CA 94305, USA}
\author[0000-0002-1200-0820]{Yao-Yuan~Mao}
\affiliation{Department of Physics and Astronomy and Pittsburgh Particle Physics, Astrophysics and Cosmology Center (PITT PACC), University of Pittsburgh, Pittsburgh, PA 15260, USA}
\author[0000-0001-5417-2260]{Gregory~M.~Green}
\affiliation{Kavli Institute for Particle Astrophysics and Cosmology and Department of Physics, Stanford University, Stanford, CA 94305, USA}
\author[0000-0003-2229-011X]{Risa~H.~Wechsler}
\affiliation{Kavli Institute for Particle Astrophysics and Cosmology and Department of Physics, Stanford University, Stanford, CA 94305, USA}
\affiliation{SLAC National Accelerator Laboratory, Menlo Park, CA 94025, USA}

\correspondingauthor{Ethan~O.~Nadler}
\email{enadler@stanford.edu}


\begin{abstract}
We develop a comprehensive and flexible model for the connection between satellite galaxies and dark matter subhalos in dark matter-only zoom-in simulations of Milky Way (MW)--mass host halos. We systematically identify the physical and numerical uncertainties in the galaxy--halo connection and simulations underlying our method, including (i) the influence of host halo properties; (ii) the relationship between satellite luminosities and subhalo properties, including the effects of reionization; (iii) the relationship between satellite and subhalo locations; (iv) the relationship between satellite sizes and subhalo properties, including the effects of tidal stripping; (v) satellite and subhalo disruption due to baryonic effects; and (vi) artificial subhalo disruption and orphan satellites. To illustrate our approach, we fit this model to the luminosity distribution of both classical MW satellites and those discovered in the Sloan Digital Sky Survey by performing realistic mock observations that depend on the luminosity, size, and distance of our predicted satellites, and we infer the total satellite population that will be probed by upcoming surveys. We argue that galaxy size and surface brightness modeling will play a key role in interpreting current and future observations, as the expected number of observable satellites depends sensitively on their surface brightness distribution.
\end{abstract}



\section{Introduction}
\label{Introduction}

Since the turn of the century, high-resolution $N$-body simulations have convincingly demonstrated that structure formation in a Lambda--Cold Dark Matter ($\Lambda$CDM) universe results in a significant number of self-gravitating DM subhalos that reside within the virial radius of Milky Way (MW)--mass host halos \citep[e.g.,][]{Klypin9901240,Moore9907411}. Relating these subhalos to observed satellite galaxies in MW-like systems requires either empirical modeling to statistically associate satellites with subhalos \citep[e.g.,][]{Brooks12095394,GarrisonKimmel13106746}, semi-analytic galaxy formation modeling (e.g., \citealt{Li09091291,Maccio09034681,Guo10060106,Starkenburg12060020,Guo150308508,Lu160502075}), or hydrodynamic simulations (e.g., \citealt{Pillepich14077855,Sawala151101098,Wetzel160205957}), all of which can yield satellite populations that are consistent with the luminosity function of the brightest MW satellites.

Modeling additional aspects of observed satellites such as their spatial, orbital, and size distributions will be necessary in order to interpret the results of current and future satellite searches in a cosmological context. Zoom-in simulations of MW-mass host halos are well-suited to this task because they provide high-fidelity realizations of the subhalo populations in these systems (e.g., \citealt{Springel08090898,GarrisonKimmel13106746,Mao150302637,Griffen150901255}). However, modeling the corresponding satellite populations using hydrodynamics is difficult because resolving MW-mass hosts along with their faint satellites requires exceptional resolution \citep{Wheeler150402466,Wetzel160205957}. In addition, sub-grid baryonic physics and star formation models, which remain uncertain, can have a significant impact on galaxy formation in this regime (\citealt{Kuhlen13055538,Munshi181012417}). While semi-analytic models offer flexible galaxy formation prescriptions that can be extended to ultra-faint systems, these approaches yield insights into the detailed astrophysical nature of the subhalo--satellite connection, rather than offering an easily interpretable coarse-grained description. The additional layer of modeling needed to track galaxy properties over time also increases the complexity of semi-analytic models relative to empirical approaches.

Predicting satellite populations directly from subhalo populations in dark matter-only (DMO) zoom-in simulations is therefore an attractive alternative. Several authors \citep[e.g.,][]{Tollerud08064381,Koposov09012116,Kravstov09063295,Hargis14074470,Kim171106267,Jethwa161207834,Newton170804247} have taken this approach to estimate the total number of MW satellites or to constrain the connection between subhalos and satellites. Many of these studies focus on specific aspects of MW satellite modeling (e.g., correcting for the completeness of observed satellite populations) and apply several distinct models to bracket the range of $\Lambda$CDM predictions.

\begin{table*}[]
\centering
\begin{tabular}{c|c|c|c}
\hline
Physical Ingredient&
Assumptions&
Parameterization&
Free Parameter?\\ \hline \hline
\multicolumn{1}{l|}{\ref{host} Host Halo Properties}& \multicolumn{1}{l|}{\hspace{7mm}Fixed by zoom-in simulations}&
\multicolumn{1}{l|}{\hspace{7mm}None}&
\multicolumn{1}{l}{\textit{\hspace{7mm}No} ($M_{\rm{host}}=10^{12.1\pm 0.03}\ \msun$)}\\ \hline

\multicolumn{1}{l|}{\ref{luminosity} Satellite Luminosities}& \multicolumn{1}{l|}{\begin{tabular}{@{}l@{}}
Abundance match to GAMA survey\\ Extrapolate luminosity function\\
Lognormal $(M_V\vert V_{\rm{peak}})$ distribution\\
No satellites below $M_{\rm{peak}}$ threshold\end{tabular}}&
\multicolumn{1}{l|}{\begin{tabular}[c]{@{}l@{}}Non-parametric\\
Faint-end slope $\alpha$\\
Constant scatter $\sigma_M$\\
Cut on $M_{\rm{peak}}<\mathcal{M}_{\rm{min}}$\end{tabular}}& \multicolumn{1}{l}{\begin{tabular}[c]{@{}l@{}}
\textit{No}\\
\textbf{Yes} ($\alpha$ is free)\\
\textbf{Yes} ($\sigma_M$ is free)\\ \textbf{Yes} ($\mathcal{M}_{\rm{min}}$ is free)\end{tabular}}\\ \hline

\multicolumn{1}{l|}{\ref{location} Satellite Locations} & \multicolumn{1}{l|}{\begin{tabular}[c]{@{}l@{}}
On-sky positions set by subhalos\\
Distances set by scaled subhalo radii\end{tabular}} & \multicolumn{1}{l|}{\begin{tabular}[c]{@{}l@{}}
None\\
$r_{\rm{sat}} \equiv \chi r_{\rm{sub}}$\end{tabular}} & \multicolumn{1}{l}{\begin{tabular}[c]{@{}l@{}}
\textit{No}\\
\textit{No} ($\chi=1$)\end{tabular}} \\ \hline

\multicolumn{1}{l|}{\ref{size} Satellite Sizes}        & \multicolumn{1}{l|}{\begin{tabular}[c]{@{}l@{}}\cite{Jiang180407306} sizes at accretion\\ Size reduction set by stripping\\Lognormal $(r_{1/2}'\vert R_{\rm{vir}})$ distribution\end{tabular}} & \multicolumn{1}{l|}{\begin{tabular}[c]{@{}l@{}}$r_{1/2}\equiv\mathcal{A}\ (c/10)^{\gamma}R_{\rm{vir}}$\\ $r_{1/2}' \equiv r_{1/2}\ (V_{\rm{max}}/V_{\rm{acc}})^{\beta}$\\ Constant scatter $\sigma_R$\end{tabular}} & \multicolumn{1}{l}{\begin{tabular}[c]{@{}l@{}}\textit{No} ($\mathcal{A}=0.02$, $\gamma=-0.7$)\\ \textit{No} ($\beta=0$)\\ \textit{No} ($\sigma_R=0.01\ \rm{dex}$)\end{tabular}} \\ \hline

\multicolumn{1}{l|}{\ref{baryonic} Baryonic Effects} & \multicolumn{1}{l|}{\hspace{7mm}\cite{Nadler171204467} disruption model} & \multicolumn{1}{l|}{\hspace{7mm}$p_{\rm{disrupt}}\rightarrow p_{\rm{disrupt}}^{1/\mathcal{B}}$} & \multicolumn{1}{l}{\hspace{7mm}\textbf{Yes} ($\mathcal{B}$ is free)} \\ \hline

\multicolumn{1}{l|}{\ref{orphan} Orphan Satellites} & \multicolumn{1}{l|}{\begin{tabular}[c]{@{}l@{}}Correspond to disrupted subhalos\\ NFW host + dynamical friction \\Stripping after pericentric passages\\ $p_{\rm{disrupt}}$ set by time since accretion\end{tabular}} & \multicolumn{1}{l|}{\begin{tabular}[c]{@{}l@{}}None \\ $\ln\Lambda = -\ln(m_{\rm{sub}}/M_{\rm{host}})$\\ $\dot{m}_{\text{sub}}\sim-\frac{m_{\rm{sub}}}{\tau_{\rm{dyn}}}\big(\frac{m_{\rm{sub}}}{M_{\rm{host}}}\big)^{0.07}$\\ $p_{\rm{disrupt}}\equiv(1-a_{\rm{acc}})^{\mathcal{O}}$\end{tabular}} & \multicolumn{1}{l}{\begin{tabular}[c]{@{}l@{}}\textit{No}\\ \textit{No} \\\textit{No}\\ \textit{No} ($\mathcal{O}=1$)\end{tabular}} \\ \hline
\end{tabular}
\caption{Summary of the physical ingredients, underlying assumptions, and parameterizations of various processes that enter our model for the subhalo--satellite connection. The final column indicates whether each component of the model is held fixed or allowed to vary for our fit to the luminosity distribution of classical and SDSS-identified satellites in Section \ref{comparison}. Bold values correspond to parameters that are varied in our fit to the observed luminosity function.}
\label{tab:master}
\end{table*}

Herein we build upon these efforts to develop a comprehensive framework that simultaneously addresses the relevant observational and modeling uncertainties. In particular, we present a flexible model for mapping subhalos to satellites and we use observations to infer the connection between these systems. Our model addresses: 

\begin{enumerate}[wide, labelwidth=!, labelindent=0pt, label=(\roman*), itemsep=0pt]
\item The influence of host halo properties on satellite populations; 
\item Satellite luminosities --- including the impact of reionization on galaxy formation --- by extrapolating an abundance matching relation to faint systems and imposing a galaxy formation threshold; 
\item The relationship between subhalo and satellite locations 
\item Satellites sizes --- including the effects of tidal stripping --- by extrapolating a galaxy size--halo size relation that accurately describes galaxy sizes in hydrodynamic simulations
\item Satellite and subhalo disruption due to baryonic effects using an algorithm calibrated on hydrodynamic zoom-in simulations of MW-mass host halos
\item Orphan satellites using a semi-analytic model to track and reinsert disrupted subhalos
\end{enumerate}

As an example application, we fit our model to the luminosity distribution of both classical MW satellites and those discovered in the Sloan Digital Sky Survey \citep[SDSS;][]{Alam150100963}, and we show that it predicts satellite populations that are qualitatively and quantitatively consistent with the luminosity function, radial distribution, and size distribution of these systems. We then forward-model the total population of MW satellites and predict satellite abundance as a function of absolute magnitude and limiting observable surface brightness; these predictions are relevant to upcoming satellite searches with improved surface brightness limits that will be carried out by surveys like the Dark Energy Survey\footnote{\https{darkenergysurvey.org}} \citep[DES;][]{DES}, the Hyper Suprime-cam Subaru Strategic Program \citep[HSC-SSP;][]{Homma170405977}\footnote{\http{hsc.mtk.nao.ac.jp/ssp}}, and the Large Synoptic Survey Telescope\footnote{\http{lsst.org}} \citep[LSST;][]{LSST}. We argue that satellite sizes, which have not consistently been included in subhalo-based models, are a key ingredient for interpreting current and future observations, as satellite detectability is highly dependent on surface brightness.

This paper is organized as follows. In Section \ref{Data}, we describe the zoom-in simulations used in this work. We present our model for connecting subhalos to satellites in Section~\ref{model}. In Section \ref{comparison}, we qualitatively and quantitatively compare our mock satellite populations to classical and SDSS-identified MW satellites. We describe an improved procedure for fitting the model to observed satellite populations, which we implement by performing mock observations and comparing our predictions to the luminosity distribution of classical and SDSS-identified systems. We discuss our results, predictions for ongoing and future surveys, implications for the low-mass subhalo--satellite connection, and caveats in Section~\ref{Results}. We summarize our model in Section~\ref{Discussion}. Throughout, we refer to bound DM systems as ``subhalos'' and to luminous galaxies as ``satellites.'' Furthermore, ``log'' refers to the base-$10$ logarithm.


\section{Simulations}
\label{Data}

We primarily use six ``MW-like'' host halos (defined below) from the suite of forty-five MW-mass DMO zoom-in simulations presented in \citet{Mao150302637}. These forty-five host halos have virial masses\footnote{We define virial quantities according to the overdensity $\Delta_{\rm vir}\simeq 99.2$ as appropriate for the cosmological parameters used in our zoom-in simulations: $h = 0.7$, $\Omega_{\rm m} = 0.286$, $\Omega_{\rm b} = 0.047$, and $\Omega_{\Lambda} = 0.714$.} in the range $M_{\rm vir}=10^{12.1\pm 0.03}\ \msun$ and have a range of formation histories that are representative of $10^{12}\ \msun$ hosts. The highest-resolution particles in these simulations have a mass of $3.0\times 10^{5}\ M_{\rm \odot}\ h^{-1}$, and the softening length in the highest-resolution regions is $170\ \text{pc}\ h^{-1}$. To test for convergence, we compare the subhalo maximum circular velocity function, radial distribution, and size distribution for one of these hosts (Halo 937) to a resimulation with a $4.0\times 10^{4}\ M_{\rm \odot}\ h^{-1}$ high-resolution particle mass and an $85\ \text{pc}\ h^{-1}$ minimum softening length. These subhalo statistics are reasonably consistent among the fiducial- and high-resolution simulations (for example, see Figure \ref{fig:orphan}), although we find a larger population of small-virial radius subhalos in the high-resolution run. 

Halo catalogs and merger trees were generated using the {\sc Rockstar} halo finder and the {\sc consistent-trees} merger code \citep{Behroozi11104372,Behroozi11104370}. \citet{Mao150302637} estimate that subhalos in these simulations are well-resolved down to a maximum circular velocity of $V_{\rm max} \approx 9\ \rm{km\ s}^{-1}$. To be conservative, we restrict our analysis to subhalos with both $V_{\rm max} > 9\ \rm{km\ s}^{-1}$ and peak maximum circular velocity $V_{\rm peak} > 10\ \rm{km\ s}^{-1}$.

The MW might be an outlier compared to typical host halos and galaxies in the relevant mass ranges (e.g., \citealt{BoylanKolchin09114484,Busha10112203,RodriguezPuebla13064328,Licquia150804446}), and its subhalo and satellite populations might be particularly influenced by the existence of the Large and Small Magellanic Clouds (LMC, SMC; e.g., \citealt{Lu160502075,Dooley170305321}). In this work, we therefore select hosts that have two Magellanic Cloud analogs, defined as subhalos with $V_{\rm{max}} > 55\ \rm{km\ s}^{-1}$ following \cite{Lu160502075}; we find six such MW-like host halos in our simulation suite. Although secondary properties of the MW in addition to the existence of the Magellanic Clouds could also bias its subhalo population \citep{Fielder180705180}, we have checked that the results presented in Section \ref{Results} are not significantly affected if we randomly select host halos from our simulation suite.


\section{Model}
\label{model}

The following subsections describe the ingredients that enter our model for the subhalo--satellite connection, which we summarize in Table \ref{tab:master}. Our model encompasses the influence of host halo properties on satellite populations (Section \ref{host}), the way in which satellites populate subhalos and the relationship between satellite and subhalo properties (Sections \ref{luminosity}--\ref{size}), and modifications to subhalo populations in DMO simulations (Sections \ref{baryonic}, \ref{orphan}). We indicate the parameters associated with our method in the subsection headers; these parameters define our physically motivated empirical model. Note that unlike in semi-analytic approaches, we do not require our model components to represent specific astrophysical processes such as star formation or quenching. In addition, although we illustrate our model using the zoom-in simulations described previously, we stress that our framework does not depend on the specific host halos used in this paper.

\subsection{Host Halo Properties}
\label{host}

Although we primarily use the six MW-like host halos described in Section~\ref{Data}, we note that our model can be applied to different hosts by changing the underlying zoom-in simulations. While the masses of our host halos are consistent with several observational constraints for the MW (e.g., \citealt{Busha10112203,Patel180301878}; also see the review in \citealt{Bland-Hawthorn160207702}), recent studies based on astrometric data suggest a more massive MW halo (\citealt{Monari180704565,Posti180501408,Simon180410230,Watkins180411348}; however, see \citealt{Callingham180810456}). Thus, studying how our predictions vary as a function of host halo mass is an important avenue for future work.

At fixed host halo mass, subhalo abundance depends on the host's formation history and secondary properties (e.g., \citealt{Zentner0411586,Fielder180705180}). For example, lower-concentration hosts accrete the majority of their subhalos later than higher-concentration hosts, leaving less time for subhalo disruption and resulting in a larger population of surviving subhalos at fixed host mass \citep{Mao150302637}. Although our MW-like host halos have lower concentration than average $10^{12}\ \msun$ hosts, marginalizing over concentration using MW-mass hosts with a range of formation histories does not affect the results of the fit presented herein.

\subsection{Satellite Luminosities}
\label{luminosity}

Next, we describe our procedure for assigning satellite luminosities to DM subhalos. We use abundance matching to link subhalos' peak maximum circular velocities $V_{\rm peak}$ to their satellites' absolute $r$-band magnitudes $M_r$ down to a certain magnitude. In particular, we follow \cite{Mao170506743} by tuning our abundance matching relation to the GAMA galaxy survey \citep{Loveday150501003} using the measured GAMA $r$-band luminosity function $k$-corrected to $z=0$ and the halo $V_{\rm{peak}}$ function from a DMO simulation with $2560^3$ particles and a side length of $250\ \text{Mpc}\ h^{-1}$, run using the same cosmological parameters and code as the Dark Sky Simulations \citep{Skillman14072600}. We abundance match to the GAMA luminosity function down to $M_r = -13\ \rm{mag}$, and we extrapolate this relation to fainter galaxies, assuming that the faint-end satellite luminosity function follows a power law. We then convert our predicted $r$-band magnitudes to $V$-band magnitudes using the empirical relation $M_r \approx M_V - 0.2\ \rm{mag}$ as in \cite{Mao170506743}. For a particular faint-end luminosity function slope, this procedure yields a mean $M_V$--$V_{\rm{peak}}$ relation. Note that by assigning magnitudes based on $V_{\rm{peak}}$, we have implicitly assumed that satellites' absolute magnitudes are not altered after accretion (see \citealt{Penarrubia07083087} for a detailed discussion of this assumption; also see \citealt{Yang11101420,Wetzel13037231,Read180707093}). We now discuss our method for varying the faint-end slope and applying scatter to our luminosity relation.

\subsubsection{Faint-end Slope ($\alpha$)}

The faint-end slope $\alpha$ of the luminosity function that enters our extrapolated abundance matching relation has been examined in previous MW satellite studies including~\cite{Tollerud08064381}. Constraints on the faint-end slope for dwarf galaxies are limited, but can in principle be derived from the total luminosity function in the Local Volume \citep{Klypin14054523} or from the satellite distributions around the MW, Andromeda (M31), or MW analogs, such as those observed by the Satellites around Galactic Analogs Survey \citep[SAGA;][]{Mao170506743}. For reference, \cite{Tollerud08064381} infer values of $-2 \lesssim \alpha \lesssim -1$ by applying completeness corrections to observed MW satellite populations. We find that a characteristic value of $\alpha=-1.3$ maps subhalos at our $V_{\rm{peak}} \approx 10\ \rm{km\ s}^{-1}$ resolution limit to satellites with $M_V \approx 5\ \rm{mag}$, which is significantly dimmer than the faintest spectroscopically confirmed MW satellite, Segue I ($M_V = -1.5\ \rm{mag}$; \citealt{McConnachie12041562}).

\begin{figure}
\centering
\includegraphics[scale=0.42]{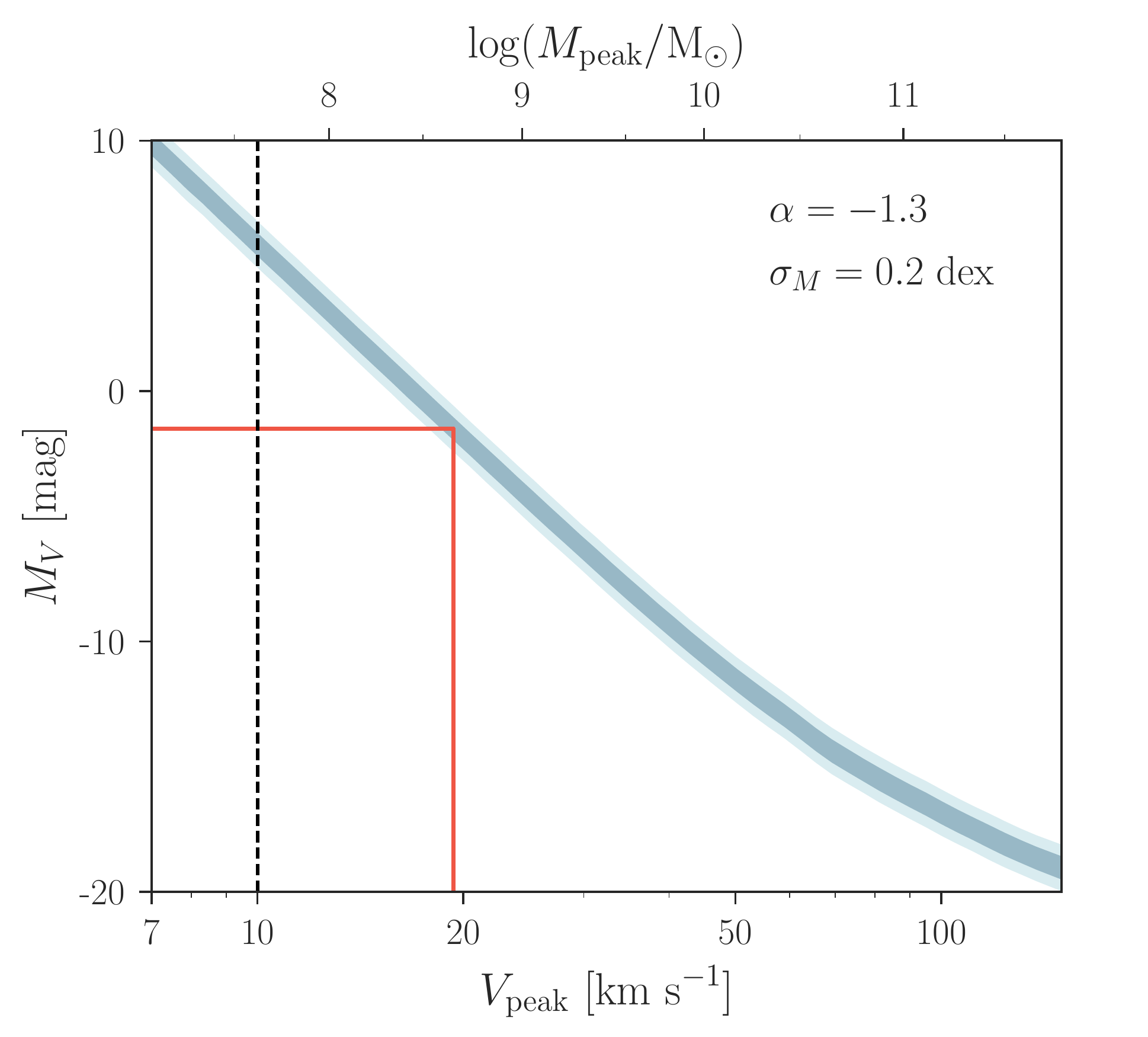}
\caption{Example of the relation between satellite luminosity and subhalo peak circular velocity used in our model. We treat the faint-end power-law luminosity function slope $\alpha$ and the constant lognormal scatter $\sigma_M$ as free parameters. Dark (light) shaded areas show $\pm 1\sigma_M$ ($\pm 2\sigma_M$) scatter. The dashed black line indicates a conservative upper bound on the resolution limit of our simulations, and the red lines indicate the absolute magnitude of the faintest spectroscopically confirmed MW satellite (Segue I; $M_V = -1.5\ \rm{mag}$) and the mean inferred $V_{\rm{peak}}$ of its subhalo for $\alpha=-1.3$ and $\sigma_M = 0.2\ \rm{dex}$.}
\label{fig:vpeakMr}
\end{figure}

\subsubsection{Luminosity Scatter ($\sigma_M$)}

Galaxy properties such as stellar mass and luminosity derived from abundance matching are usually assumed to follow lognormal distributions at fixed halo mass proxy, and the scatter in these properties is relatively well constrained to $\sim 0.2\ \rm{dex}$ for host halos with masses above $\sim 10^{12}\ \msun$ (\citealt{Behroozi12076105,Reddick12072160,Lehmann151005651}; also see the review in \citealt{Wechsler180403097}).\footnote{The observational uncertainties associated with the conversion from stellar mass to luminosity are likely larger than the difference in scatter between these quantities \citep{Wechsler180403097}, so we ignore this distinction here.} However, this scatter is not well constrained in the MW satellite regime \citep{Garrison-Kimmel160304855,Munshi170506286}. We therefore treat the scatter in our predicted absolute magnitudes at fixed $V_{\rm{peak}}$, which we denote $\sigma_M$, as a free parameter with a lower bound of $0.2\ \rm{dex}$. In particular, we apply scatter to each satellite's absolute magnitude by drawing from a lognormal distribution with a mean set by our $M_V$--$V_{\rm{peak}}$ relation and a standard deviation of $\sigma_M$. Note that we do not deconvolve the abundance matching relation when we vary the scatter for computational efficiency; we have checked that this choice does not significantly affect the resulting $M_V$--$V_{\rm{peak}}$ relation. Although several semi-analytic models (e.g., \citealt{Guo150308508}) and hydrodynamic simulations (e.g., \citealt{Sawala14066362,Fitts161102281}) suggest that the scatter between galaxy properties and halo properties grows with decreasing halo mass, we treat $\sigma_M$ as a constant for the fit presented in this paper. It might be necessary to relax this assumption in future work that addresses a more complete population of observed satellites.

Because we extrapolate our abundance matching relation to systems fainter than $M_r = -13\ \rm{mag}$, our procedure is not designed to match the global luminosity function in this regime; however, the global constraint is not our primary concern, and our power-law-plus-scatter parameterization results in a more flexible model. We show an example $M_V$--$V_{\rm{peak}}$ relation in Figure~\ref{fig:vpeakMr} for fiducial choices of $\alpha$ and $\sigma_M$, but we emphasize that these are free parameters in our model. Figure~\ref{fig:vpeakMr} illustrates that our simulations should resolve all subhalos that host satellites with $M_V < 0\ \rm{mag}$ for this choice of $\alpha$ and $\sigma_M$ (however, see the discussion on artificial subhalo disruption and orphan satellites in Section \ref{orphan}).

\subsubsection{Galaxy Formation Threshold ($\mathcal{M}_{\rm{min}}$)}

Many authors have studied the impact of reionization on galaxy formation in low-mass subhalos, finding that subhalos below a certain mass or $V_{\rm{peak}}$ threshold are likely to be dark (e.g., \citealt{Thoul9510154,Bullock0002214,Somerville0107507}). Because we simply assign satellite luminosities to subhalos, we must account for this effect; however, the details of the galaxy formation process for faint satellites are unclear. For example, \cite{Kuhlen13055538}, \cite{Sawala14043724}, and \cite{Fitts161102281} respectively find that most isolated halos with $M_{\rm{peak}}\approx 10^{10}\ \msun$ ($V_{\rm{peak}}\approx 45\ \rm{km\ s}^{-1}$), $3\times 10^{9}\ \msun$ ($V_{\rm{peak}}\approx 33\ \rm{km\ s}^{-1}$), and $10^{9}\ \msun$ ($V_{\rm{peak}}\approx 25\ \rm{km\ s}^{-1}$) are dark. Meanwhile, \cite{Jethwa161207834} find that the peak virial mass of the subhalo hosting Segue I ($M_V = -1.5\ \rm{mag}$) is below $2.4 \times 10^{8}\ \msun$ ($V_{\rm{peak}}\approx 16\ \rm{km\ s}^{-1}$) at the $68\%$ confidence level, and \cite{Bland-Hawthorn150506209} find that star formation can proceed after supernova feedback in subhalos with peak virial masses down to $\sim 10^{7}\ \msun$ ($V_{\rm{peak}}\approx 7\ \rm{km\ s}^{-1}$).

These results are generally sensitive to the assumed redshift of reionization and ultraviolet background; to account for these uncertainties in a simple way, we treat the minimum peak virial mass necessary for galaxy formation, $\mathcal{M}_{\rm{min}}$, as a free parameter. In particular, for a given value of $\mathcal{M}_{\rm{min}}$, we discard satellites in all subhalos with $M_{\rm{peak}}<\mathcal{M}_{\rm{min}}$. This mass cut clearly does not capture the complexities of galaxy formation in low-mass subhalos, which likely result in a smoothly varying galaxy occupation fraction rather than a sharp cutoff \citep{Sawala14043724,Sawala14066362,Fitts180106187}. Although it is not \emph{necessary} to introduce $\mathcal{M}_{\rm{min}}$ because we fit to an incomplete sample of observed satellites, we will demonstrate that the classical-plus-SDSS luminosity distribution sets an interesting upper bound on this quantity, given our simple parameterization.

\subsection{Satellite Locations}
\label{location}

Next, we describe our procedure for assigning on-sky coordinates and radial distances to our mock satellites.

\subsubsection{On-sky Positions}

In general, we expect the positions of satellite galaxies to correspond reasonably well to the positions of their subhalos. Thus, we simply use the projected on-sky positions of subhalos in our zoom-in simulations to assign on-sky coordinates to our satellites. Since our DMO simulations do not contain galactic disks, we are free to perform arbitrary 3D rotations of our subhalo positions about host halo centers before projecting them onto the sky; these rotations can be fixed based on the positions of Magellanic Cloud--like subhalos in order to perform realistic mock MW satellite surveys. In addition, we convert satellites' Galactocentric coordinates to heliocentric coordinates by placing mock observers $8\ \rm{kpc}$ from our host halo centers, as described in Section \ref{comparison}.

\subsubsection{Radial Scaling ($\chi$)}

MW satellites seem to be unusually centrally concentrated compared to both the observed satellite population in M31 and typical subhalo populations in zoom-in simulations (e.g., \citealt{Yniguez13050560,Graus180803654}; however, see \citealt{Li180600041}). While this apparent discrepancy might be caused in part by misestimates of observational incompleteness, several numerical effects could contribute to a mismatch between the radial distribution of simulated subhalos and observed satellites. For example, simulations might underestimate the amount of dynamical friction experienced by subhalos due to resolution effects (although we expect this to be a subdominant source of error in MW-mass systems), and halo finders might mis-track or fail to identify subhalos in dense, central regions \citep{Knebe11040949}. We expect halo finder incompleteness to be mitigated in our analysis, given that we use the phase-space-based halo finder {\sc Rockstar} and because we reinsert disrupted subhalos using the orphan model described in Section \ref{orphan}.

To model these potential biases in our radial satellite distributions, we define the parameter $\chi\in (0,1]$ as follows:
\begin{equation}
r_{\rm{sat}} \equiv \chi r_{\rm{sub}},
\end{equation}
where $r_{\rm{sat}}$ is a satellite's distance from the center of its host halo (which we identify with its Galactocentric distance) and $r_{\rm{sub}}$ is the Galactocentric distance of the corresponding subhalo. Thus, $\chi=1$ corresponds to setting each satellite's radial distance equal to that of its subhalo, while smaller values of $\chi$ shift our mock satellites inward relative to their subhalos. Although we fix $\chi=1$ for the fit to the luminosity distribution of classical and SDSS-identified satellites presented below, we include it in our modeling framework for generality and as a useful phenomenological parameter for future work.

\subsection{Satellite Sizes}
\label{size}

To assign sizes to our mock satellites, we use a modified version of the galaxy size--halo virial radius relation from \cite{Jiang180407306}, which relates a galaxy's 3D half-mass radius to its subhalo's virial radius $R_{\rm{vir}}$. Since we will compare the sizes of our mock satellites to measured half-light radii, we simply identify our satellites' predicted 3D half-mass radii with their projected 2D half-light radii. This conversion neglects mass-to-light weighting and projection effects; these are both reasonable approximations, although the latter overestimates the sizes of highly elliptical dwarfs.

\subsubsection{Mean Size Relation ($\mathcal{A},\gamma$)}

We use the following relation from \cite{Jiang180407306} to set satellite half-light radii at the time of accretion:
\begin{equation}
r_{1/2} \equiv \mathcal{A}\ \Big(\frac{c}{10}\Big)^\gamma\ R_{\rm{vir}},\label{eq:size}
\end{equation}
where $\mathcal{A} = 0.02$, $\gamma = -0.7$, and $c$ denotes subhalo concentration measured at accretion.\footnote{We have also tested the size model from \cite{Kravstov12122980}, but we find that the concentration dependence in the \cite{Jiang180407306} relation leads to a more reasonable range of sizes compared to observed MW satellites.} 
\cite{Jiang180407306} find that this relation yields galaxy sizes that are consistent with those found in in two hydrodynamic simulations with a residual scatter of $\sim 0.15\ \rm {dex}$ about the hydrodynamic results. Although Equation \ref{eq:size} is essentially untested for ultra-faint dwarf galaxies, we hold $\mathcal{A}$ and $\gamma$ fixed at their fiducial values, and we account for this uncertainty by allowing the amount of size reduction that satellites undergo and the scatter in this relation to vary as described in the following subsections.

\subsubsection{Size Reduction Due to Tidal Stripping ($\beta$)}

One might expect the sizes of satellite galaxies to correlate tightly with their subhalos' virial radii at accretion. However, dynamical effects such as tidal stripping can alter subhalo and satellite sizes after accretion, introducing scatter in this relationship (e.g., \citealt{Kravstov0401088,Penarrubia07083087}). To model these effects, we calculate each satellite's $z=0$ half-light radius $r_{1/2}'$ as follows:
\begin{equation}
r_{1/2}' \equiv r_{1/2}\ \Big(\frac{V_{\rm{max}}}{V_{\rm{acc}}}\Big)^{\beta},\label{eq:size2}
\end{equation}
where $V_{\rm{max}}$ denotes the corresponding subhalo's maximum circular velocity at $z=0$, $V_{\rm{acc}}$ denotes its maximum circular velocity at accretion, and $\beta \geq 0$ is a model parameter. Thus, $\beta = 0$ corresponds to no change in satellite sizes after accretion, while larger values of $\beta$ increase the amount by which satellite sizes are reduced based on the degree of tidal stripping that their subhalos undergo. This prescription does not capture the effects of tidal heating, which can \emph{enlarge} satellites rather than reduce their sizes. In addition, although $V_{\rm{max}}/V_{\rm{acc}}$ depends on the orbital history of each subhalo, the amount of tidal stripping that their satellites undergo might also depend on orbital inclination with respect to the central disk, an effect that is not modeled in our DMO simulations. Thus, Equation \ref{eq:size2} will likely need to be recalibrated on the size evolution of ultra-faint satellites in upcoming hydrodynamic simulations. Pending such results and a detailed model for stellar mass stripping, we fix $\beta=0$ for our fiducial model, and we demonstrate in Appendix \ref{appendix} that our constraints are insensitive to this parameter.

\begin{figure}
\centering
\includegraphics[scale=0.425]
{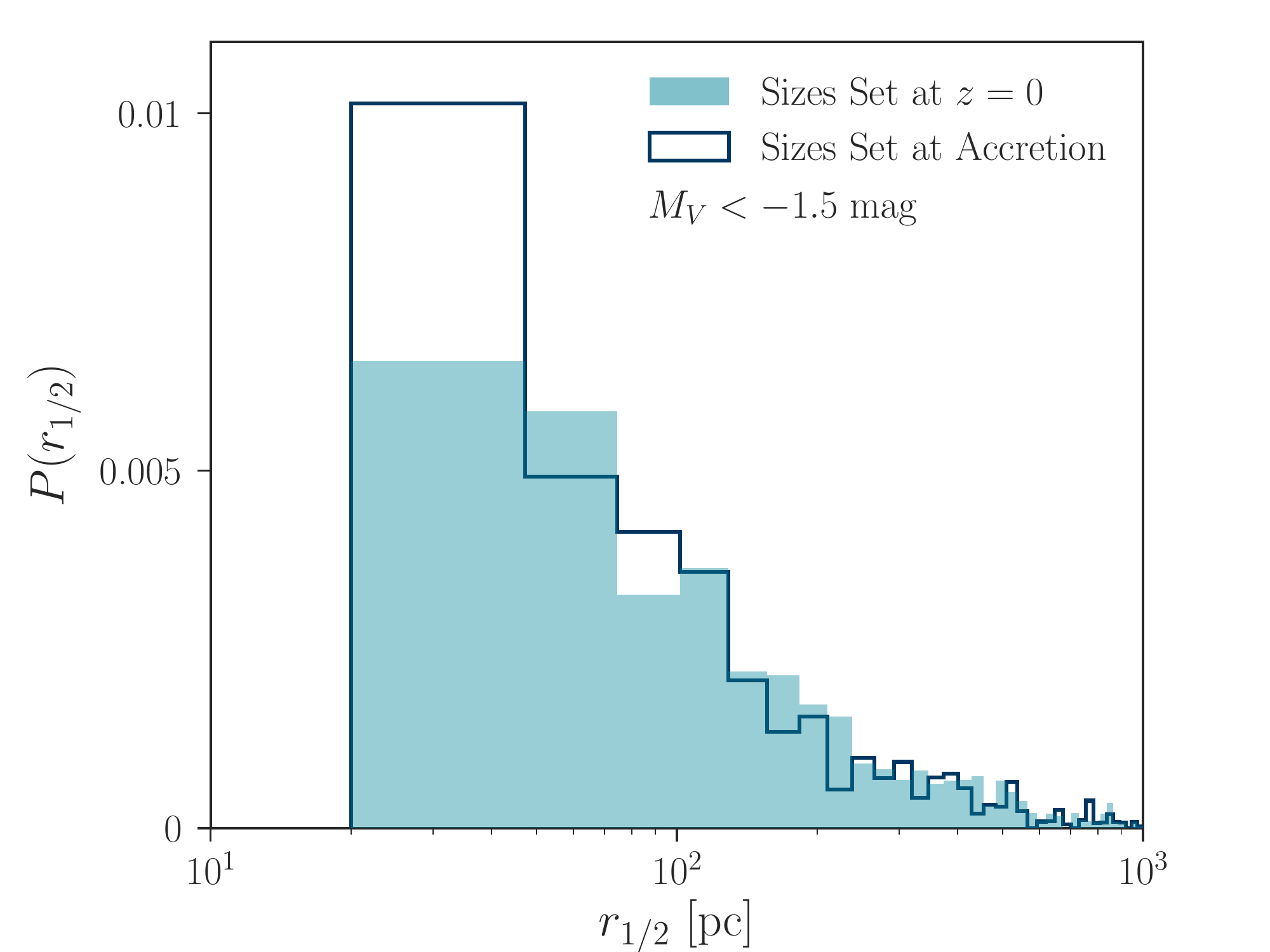}
\caption{Size distributions for mock satellites in our six MW-like host halos, with satellite sizes set by subhalo sizes at accretion ($\beta=0$) and at $z=0$ using our prescription for satellite size reduction due to tidal stripping ($\beta=1$). The distributions are weighted by survival probability using our subhalo disruption model with $\mathcal{B}=1$.}
\label{fig:sizehist}
\end{figure}

Figure \ref{fig:sizehist} compares the mean size distribution at accretion ($\beta=0$) and at $z=0$ using our tidal stripping model ($\beta=1$) for mock satellites in our six MW-like host halos. Satellite sizes are reduced due to tidal stripping, as expected from Equation~\ref{eq:size2}. Our extrapolation of the \cite{Jiang180407306} relation yields a reasonable range of sizes compared to observed MW satellites, which roughly span $10\ \text{pc}\lesssim r_{1/2} \lesssim 3000\ \rm{pc}$.

\subsubsection{Size Scatter ($\sigma_{R}$)}

The size relation that we have described has not been tested against hydrodynamic simulations for galaxies with half-light radii smaller than $\sim 400\ \rm{pc}$. Thus, for small systems, the uncertainty in this relation might deviate from the $\sim 0.15\ \rm{dex}$ residuals found in \cite{Jiang180407306}. We therefore apply scatter by drawing each satellite's size from a lognormal distribution with a mean given by Equation \ref{eq:size2} and a standard deviation of $\sigma_R$, and we impose a minimum size of $20\ \rm{pc}$ as appropriate for the classical and SDSS-identified satellites studied herein. Although this scatter might be size-dependent, we treat $\sigma_R$ as a constant for simplicity.

\begin{figure*}
\centering
\includegraphics[scale=0.4]{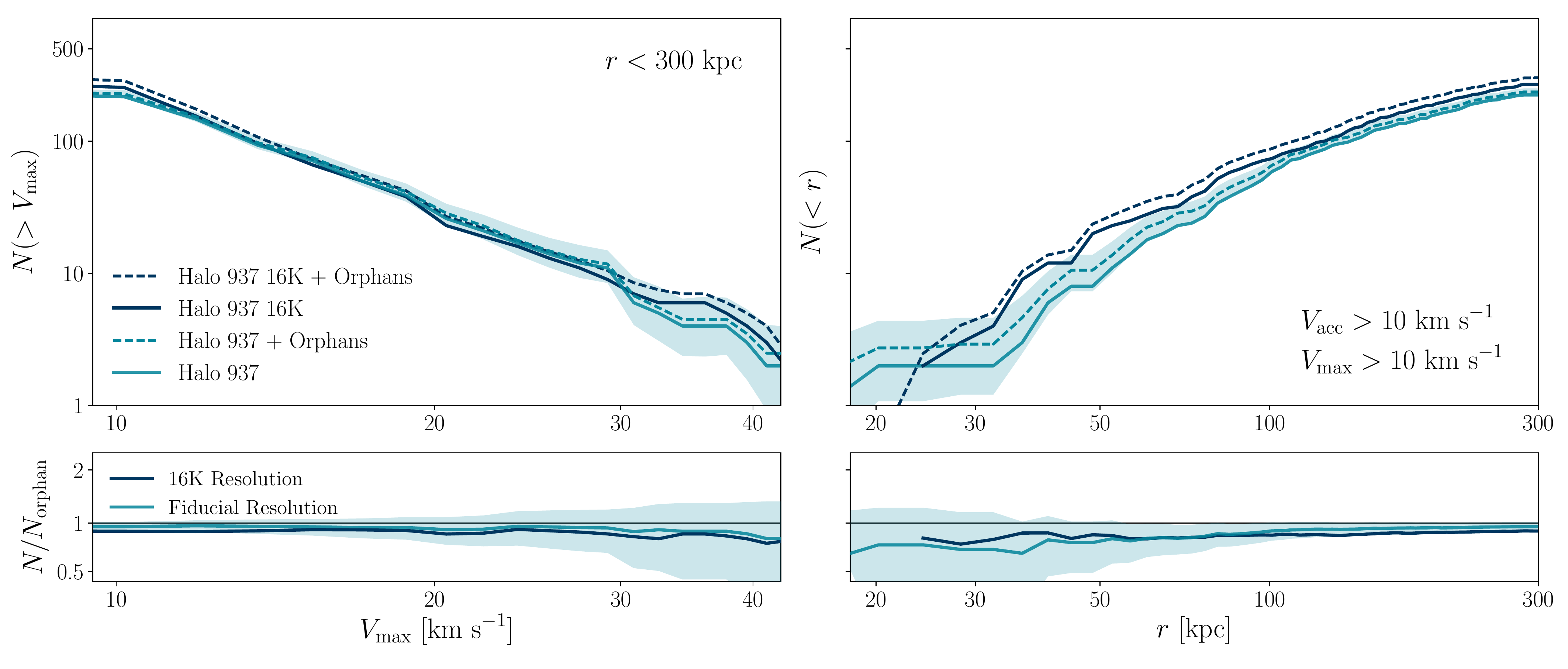}
\caption{$V_{\rm{max}}$ function (left panel) and radial distribution (right panel) of subhalos with $V_{\rm acc}>10\ \rm{km\ s}^{-1}$ and $V_{\rm max}>10\ \rm{km\ s}^{-1}$ in one of our zoom-in simulations, shown with (dashed lines) and without (solid lines) our fiducial orphan model ($\mathcal{O}=1$). The dark blue lines show results from a high-resolution simulation of this host halo, and the blue bands show the Poisson scatter for our prediction that includes all disrupted subhalos tracked to $z=0$ in the fiducial-resolution simulation. The bottom panels show the ratio of the number of subhalos to the number of subhalos in the corresponding simulation including orphans for our fiducial-resolution (blue) and high-resolution (dark blue) runs. Including orphans brings the fiducial-resolution $V_{\rm{max}}$ function and radial distribution into fairly good agreement with the high-resolution results.}
\label{fig:orphan}
\end{figure*}

\subsection{Subhalo Disruption due to Baryonic Effects ($\mathcal{B}$)}
\label{baryonic}

To model the enhancement in subhalo disruption due to baryonic effects such as the presence of a central galactic disk, we apply the subhalo disruption algorithm from \cite{Nadler171204467}, which uses the orbital and internal properties of subhalos in DMO simulations to predict the probability that they will be disrupted in hydrodynamic resimulations. This model was trained on two hydrodynamic simulations of MW-mass host halos from the Feedback In Realistic Environments project (FIRE; \citealt{Hopkins13112073,Hopkins170206148}), both of which have classical satellite populations that are reasonably consistent with those in the MW and M31 \citep{Wetzel160205957,Garrison-Kimmel170103792}. \cite{Nadler171204467} found that this disruption model predicts surviving subhalo populations that are in better agreement with hydrodynamic results than DMO simulations that include the gravitational potential of the galactic disk found in corresponding hydrodynamic simulations. Our method therefore captures both the tidal influence of a central disk and additional baryonic effects, so our work differs from studies that employ DM-plus-disk simulations to model the effects of baryons on subhalo populations (e.g., \citealt{Jethwa161207834,Newton170804247}).

\cite{Nadler171204467} showed that baryonic subhalo disruption approximately rescales the subhalo $V_{\rm{max}}$ functions for the zoom-in simulations used in this work. However, the radial subhalo distributions are not scaled by a constant factor due to enhanced subhalo disruption in the inner $\sim 50\ \rm{kpc}$ of host halos caused by the dynamical influence of the central galactic disks in the FIRE simulations. To account for the uncertainty associated with the limited training set in \cite{Nadler171204467} --- and particularly the limited set of host halo \emph{accretion histories} that the model is trained on --- we define the parameter $\mathcal{B}$ to characterize subhalo disruption due to baryonic effects as follows. We assign each surviving subhalo in our DMO simulations a disruption probability by modifying the prediction from the \cite{Nadler171204467} model according to $p_{\rm disrupt} \rightarrow p_{\rm disrupt}^{1/\mathcal{B}}$. Thus, $\mathcal{B} = 1$ corresponds to the unaltered disruption probabilities, $\mathcal{B}>1$ ($\mathcal{B}<1$) corresponds to increased (decreased) disruption probabilities relative to our fiducial model, and $\mathcal{B} = 0$ corresponds to no additional subhalo disruption due to baryonic effects. We assume that each satellite's disruption probability is equal to that of its subhalo. Although this assumption is reasonable for disruption mechanisms like disk shocking, it warrants further investigation using controlled simulations.

\subsection{Orphan Satellites ($\mathcal{O}$)}
\label{orphan}

Cosmological simulations often require a population of orphan galaxies (i.e., galaxies whose halos are not detected by the halo finder) to match observational galaxy clustering constraints (e.g., \citealt{Wang0603546,Guo13033586,Pujol170202620}). The details of orphan modeling depend on the simulation and target galaxy population in question; however, orphans should generally be included in analyses that are sensitive to systems near a resolution threshold. Thus, despite the relatively high resolution of our zoom-in simulations, orphans are potentially important in our study, since we aim to model faint MW satellites. Moreover, if \emph{artificial} subhalo disruption is a significant effect (e.g., \citealt{VandenBosch180105427,VandenBosch171105276}), in the sense that disrupted subhalos in our simulations should host observable satellite galaxies, then it becomes even more important to include orphans.

To model orphans, we therefore identify all disrupted subhalos in each simulation that contribute directly to the main host halo. 
We track the orbit of each disrupted subhalo until $z=0$ using a softened gravitational force law and dynamical friction as follows:
\begin{equation}
\mathbf{\dot{v}} = -\frac{GM(<r)}{(r+\epsilon R_{\rm vir,host})^2}\mathbf{\hat{r}} + \frac{\mathbf{F}_{\rm{df}}}{m_{\rm{sub}}}.
\end{equation}
Here, $r$ is a subhalo's distance to the center of the host, $m_{\rm{sub}}$ is its virial mass, $M(<r)$ is the enclosed host halo mass\footnote{We calculate $M(<r)$ assuming an NFW host halo density profile---that is, 
$M(<r) = M_{\rm{host}}f(cx)/f(c)$, where $x\equiv r/R_{\rm{vir,host}}$, $c$ is the host halo's concentration, and $f(\xi)\equiv \ln(1+\xi) - \xi/(1+\xi)$.}, $R_{\rm{vir,host}}$ is the host halo's virial radius, and the gravitational softening $\epsilon=0.01$ is chosen to avoid hard collisions with the host (we have checked that the resulting subhalo orbits are insensitive to this choice for reasonably small values of $\epsilon$). To calculate $\mathbf{F}_{\rm{df}}$, we use the \cite{Chandrasekhar1943} dynamical friction formula for an NFW host halo and a Maxwellian distribution of host particle velocities, which yields
\begin{equation}
\mathbf{F}_{\rm{df}} = -4\pi \Big(\frac{Gm_{\rm{sub}}}{\mathopen|\mathbf{v}_{\rm{orb}}\mathclose|}\Big)^2\ \ln \Lambda\ \rho(r)\Big[\text{erf}(X) - \frac{2X}{\sqrt{\pi}}e^{-X^2}\Big]\frac{\mathbf{v}_{\rm{orb}}}{\mathopen|\mathbf{v}_{\rm{orb}}\mathclose|},
\end{equation}
where $\mathbf{v}_{\rm{orb}}$ denotes subhalo orbital velocity, $\ln \Lambda$ is the Coulomb logarithm, $\rho(r)$ is the host halo's density profile, and $X \equiv \text{v}_{\text{orb}}/[\sqrt{2}\sigma(r)]$ where $\sigma(r)$ is the local host halo velocity dispersion. We estimate $\sigma(r)$ using the fitting formula in \cite{Zentner0304292} and we set $\ln \Lambda = -\ln (m_{\rm{sub}}/M_{\rm{host}})$ following \cite{Gan10070023}.

To account for tidal stripping, we follow \cite{Behroozi180607893} by modeling mass loss for disrupted subhalos as follows:
\begin{eqnarray}
\dot{m}_{\rm sub,infalling} &=& 0\\
\dot{m}_{\rm sub,outgoing} &=& -1.18\ \frac{m_{\rm{sub}}}{\tau_{\rm dyn}}\Big(\frac{m_{\rm{sub}}}{M_{\rm{host}}}\Big)^{0.07},
\end{eqnarray}
where $\tau_{\rm dyn} = (4\pi G\rho_{\rm vir}/3)^{-1/2}$ is the dynamical timescale and derivatives are taken with respect to time. This mass stripping model is motivated by several synthetic and cosmological tests that have shown that the majority of subhalo mass loss occurs after pericentric passages \citep{Knebe11040949,Behroozi13102239}. We use a modified version of the fitting formula from \cite{Jiang4582848} to model the corresponding reduction in each disrupted subhalo's maximum circular velocity,
\begin{equation}
\frac{\text{d}\log V_{\rm max}}{\text{d}\log m_{\rm{sub}}} = 0.3 - 0.4\frac{m_{\rm{sub}}}{m_{\rm{sub}}+m_{\rm sub,acc}},
\end{equation}
where $m_{\rm sub,acc}$ denotes subhalo virial mass at accretion. Finally, we calculate the sizes of our orphan satellites using Equations~\ref{eq:size}--\ref{eq:size2}.

To vary the contribution from orphan satellites, we define the parameter $\mathcal{O}$ by setting the disruption probability for each orphan equal to $(1 - a_{\rm acc})^{\mathcal{O}}$, where $a_{\rm{acc}}$ is the final scale factor at which a subhalo enters the virial radius of its host halo. We find that this formula with $\mathcal{O}=1$ describes the disruption probabilities predicted by the \cite{Nadler171204467} model for \emph{surviving} subhalos fairly well, and we use it because the uncertainties in the other features that enter the \cite{Nadler171204467} model (e.g., pericentric distance) are potentially large for disrupted subhalos. Thus, $\mathcal{O} = 0$ corresponds to including zero orphan satellites, larger values of $\mathcal{O}$ increase the contribution from orphans, and each orphan's disruption probability grows with the amount of time elapsed since accretion.

To test our orbit tracking and tidal stripping models, we calculate $V_{\rm{max}}$ functions and radial subhalo distributions for the fiducial- and high-resolution versions of Halo 937 described in Section \ref{Data} by selecting subhalos with $V_{\rm{acc}}>10\ \rm{km\ s}^{-1}$ and $V_{\rm{max}} > 10\ \rm{km\ s}^{-1}$ (recall that $V_{\rm{max}}\approx 9\ \rm{km\ s}^{-1}$ corresponds to the resolution threshold of our fiducial simulations). Figure~\ref{fig:orphan} shows that our orphan model with $\mathcal{O}=1$ brings both the velocity function and radial distribution from the fiducial-resolution simulation (light blue dashed lines) into fairly good agreement with the high-resolution results (dark blue solid lines). While the change in the velocity function due to orphans is consistent with a change in overall normalization (i.e., a constant scaling with respect to $V_{\rm{max}}$) at both resolution levels, we find that orphans are preferentially added in the inner regions of our fiducial-resolution simulation, where they are most likely to be needed. On the other hand, our orphan model merely rescales the radial subhalo distribution in the high-resolution run, suggesting that there is less spurious disruption in this case. 

We note that our orphan model can be interpreted in terms of the amount of artificial subhalo disruption that occurs in our simulations (e.g., \citealt{VandenBosch180105427,VandenBosch171105276}). In particular, $\mathcal{O}=0$ (no orphans included) corresponds to the assumption that all subhalo disruption in our simulations is both physical and coincides with the disruption of their satellites. Meanwhile, $\mathcal{O}\gg 1$ (all orphans included with zero disruption probability) implies that all subhalo disruption in our DMO simulations and all \emph{additional} subhalo disruption due to baryonic effects in hydrodynamic simulations is artificial, in the sense that all disrupted subhalos should host satellites. We find that these extreme possibilities are respectively disfavored by the inner radial distribution and the total abundance of observed MW satellites, which effectively set lower and upper bounds on $\mathcal{O}$. We fix $\mathcal{O}=1$ for the fit presented as follows, which roughly corresponds to the assumption that subhalo disruption in our DMO simulations is artificial while subhalo disruption in the hydrodynamic simulations that we train our disruption model on is physical. This assumption does not necessarily contradict the results of \cite{VandenBosch171105276} and \cite{VandenBosch180105427}, since their subhalo disruption tests were performed without central disk potentials. Our hope is that future data will better constrain this parameter.

\subsection{Comparison to Recent Models}

Our approach differs from previous models for the subhalo--satellite connection in several regards. To illustrate these differences, we compare our model to those recently presented in \cite{Jethwa161207834} and \cite{Newton170804247}:
\begin{enumerate}
\item Our procedure for assigning satellite luminosities to subhalos is tuned to match an observed luminosity function for systems brighter than $M_r = -13\ \rm{mag}$, unlike the empirical stellar mass--halo mass relations considered in \cite{Jethwa161207834}; meanwhile, \cite{Newton170804247} estimate the total number of MW satellites by statistically comparing radial subhalo distributions in the {\sc Aquarius} simulations to classical and SDSS-identified satellites without explicitly modeling the luminosity distribution of these systems.
\item We model the sizes of our mock satellites, while recent empirical studies, including \cite{Jethwa161207834} and \cite{Newton170804247}, assume that all systems of a given luminosity have sufficient surface brightness to be observed (though see \citealt{Bullock09121873}).
\item Our model for baryonic subhalo disruption is similar to the prescriptions in \cite{Jethwa161207834} and \cite{Newton170804247}, which are based on DM-plus-disk simulations. However, our algorithm predicts surviving subhalo populations that are in better agreement with hydrodynamic results compared to DM-plus-disk simulations for the host halos that it was trained on \citep{Nadler171204467}.
\item We parameterize our baryonic disruption and orphan satellite models to allow for deviations from our fiducial prescriptions, unlike \cite{Jethwa161207834} and \cite{Newton170804247}.
\end{enumerate}


\section{Comparison to Observed Satellites}
\label{comparison}

We now demonstrate that our model can produce satellite populations that are both qualitatively and quantitatively consistent with classical and SDSS-identified MW satellites. We reiterate that other satellite populations --- including those associated with M31 or with MW-mass hosts outside of the Local Volume --- can be used to constrain our model, but we focus on the well-characterized population of classical and SDSS-identified MW satellites for clarity. Thus, we \emph{do not} utilize all known MW satellites for the fit presented as follows.

\subsection{Qualitative Comparison}
\label{visualization}

\begin{figure*}
\centering
\includegraphics[scale=0.46]{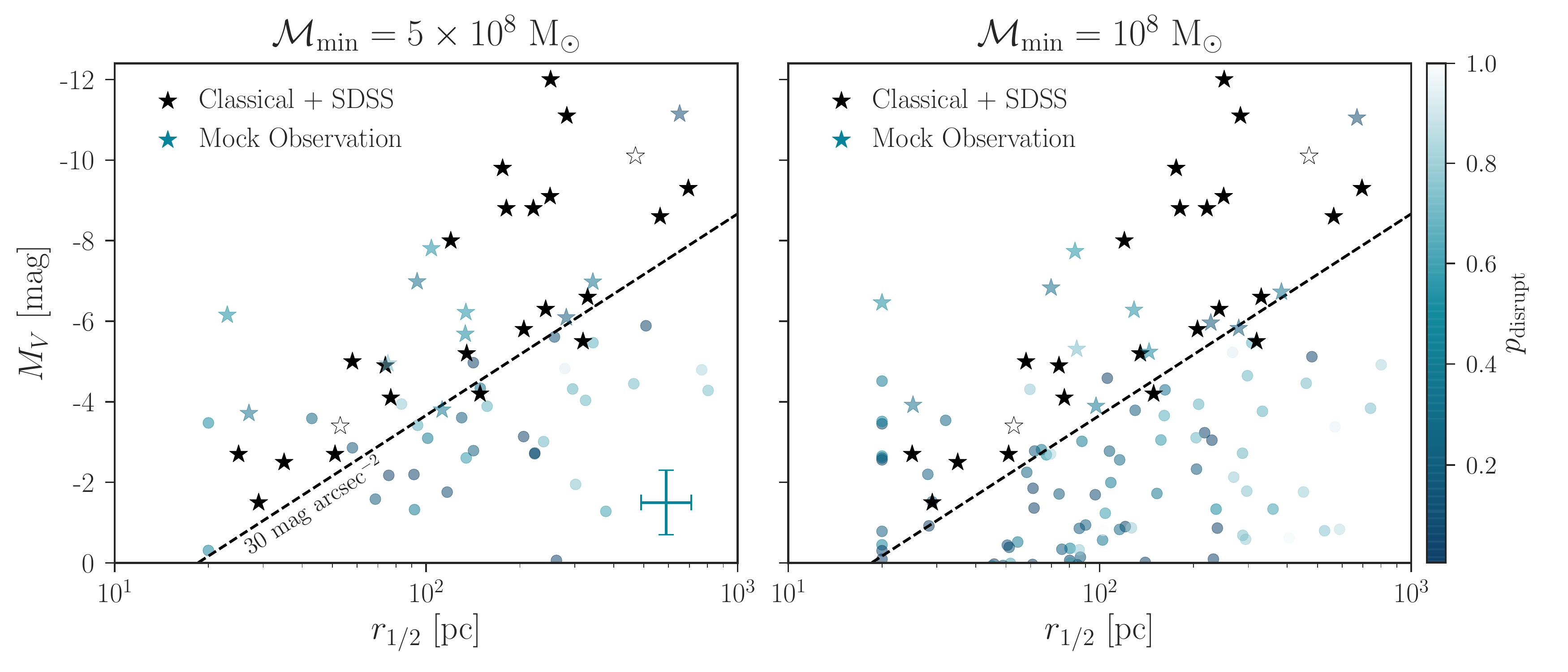}
\includegraphics[scale=0.46]{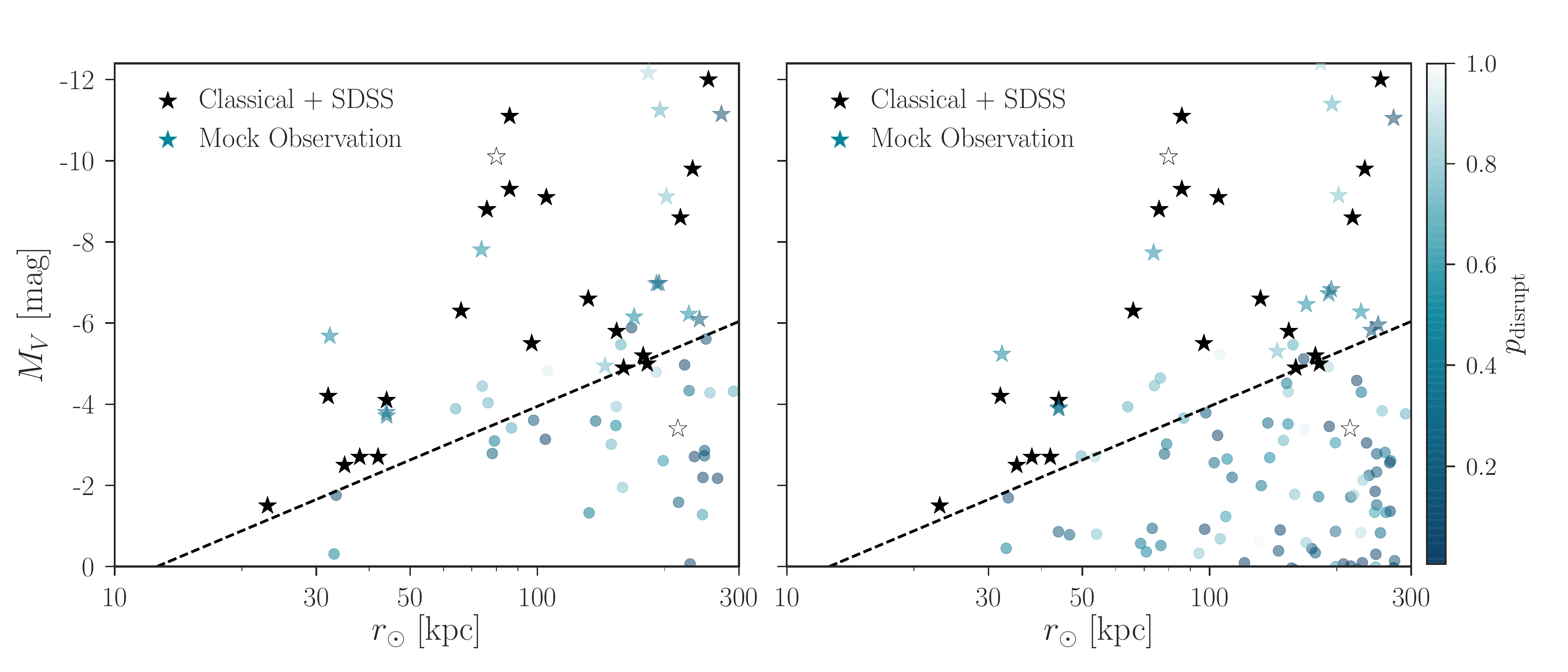}
\caption{Visualizations of the predicted satellite population in one of our MW-like host halos (blue symbols) and the observed population of classical and SDSS-identified MW satellites (black stars) in the absolute magnitude--half-light radius plane (top panels) and the absolute magnitude--heliocentric distance plane (bottom panels) for $\mathcal{M}_{\rm{min}}=5\times 10^{8}\ \msun$ (left panels) and $\mathcal{M}_{\rm{min}}=10^{8}\ \msun$ (right panels). All classical satellite analogs ($M_V\leqslant -8.8\ \rm{mag}$) and all systems in a mock SDSS footprint that pass both the SDSS surface brightness and completeness radius detection limits \citep[dashed lines;][]{Koposov07062687,Koposov09012116} are plotted as blue stars, while blue circles show systems in the mock SDSS footprint that do not pass both detection criteria. The color bar indicates satellite disruption probability. This realization uses a faint-end slope of $\alpha = -1.3$, a luminosity scatter of $\sigma_M=0.2\ \rm{dex}$, and a disruption parameter of $\mathcal{B} = 1$; we fix the remaining parameters according to Table \ref{tab:master}. The error bars in the top-left panel show characteristic uncertainties for $M_V$ and $r_{1/2}$. Pisces I and Pegasus III, which were discovered in SDSS using methods that do not adhere to our assumed detection criteria, are plotted as unfilled stars.}
\label{fig:vis}
\end{figure*}

Before fitting our model to observed MW satellites, we qualitatively compare its predictions to the abundance and properties of classical and SDSS-identified systems. In particular, Figure~\ref{fig:vis} shows projections of the predicted satellite population for one of our MW-like host halos in the observationally motivated parameter space of absolute magnitude, half-light radius, and heliocentric distance using fiducial values of our free parameters. We compare our predictions to the following classical and SDSS-identified MW satellites compiled in \cite{McConnachie12041562}: LMC, SMC, Sagittarius I, Fornax, Leo I, Sculptor, Leo II, Sextans I, Carina, Draco, and Ursa Minor (classical), and Canes Venatici I, Hercules, Bo{\"o}tes I, Leo IV, Ursa Major I, Leo V, Pisces II, Canes Venatici II, Ursa Major II, Coma Berenices, Willman I, Bo{\"o}tes II, Segue II, and Segue I (SDSS). We exclude Pisces I and Pegasus III because they were discovered using methods that do not adhere to our assumed SDSS detection criteria (described as follows), and we exclude Leo T because it lies outside of our fiducial $300\ \rm{kpc}$ reference radius. We refer the reader to \cite{McConnachie12041562} for references to the papers in which these systems were discovered.

Figure \ref{fig:vis} illustrates all mock classical satellites in one of our host halos, along with all systems in a region corresponding to the area of the SDSS survey that points away from the LMC analog in this simulation. We define mock classical satellites as objects with $M_V \leqslant -8.8\ \rm{mag}$ and we assume that observations of these systems are complete. We plot systems within $300\ \rm{kpc}$ that pass both the surface brightness limit of $30\ \rm{mag\ arcsec}^{-2}$ estimated in \cite{Koposov07062687} and the distance--magnitude SDSS detection limit estimated in \cite{Koposov09012116} as blue stars, while blue circles indicate systems in the mock SDSS footprint that \emph{do not} pass both detection criteria. The \cite{Koposov09012116} heliocentric completeness radius can be expressed as
\begin{equation}
R_{\rm{eff}}(M_V) = 10^{-aM_V+b}\ \rm{kpc},\label{eq:walsh}
\end{equation}
where $a = 0.228$ and $b=1.1$. Thus, a satellite with magnitude $M_V$ passes this detection criterion if it falls within the effective radius $R_{\rm{eff}}(M_V)$ given by Equation \ref{eq:walsh}. Adopting the \cite{Walsh08073345} version of the SDSS detection threshold, which corresponds to $a=0.187$ and $b=1.42$, yields similar constraints for the fit presented herein. For each mock satellite, we calculate absolute surface brightness using the relation
\begin{equation}
\mu_V = M_V + 36.57 + 2.5\log[2\pi(r_{1/2}'/1\ \rm{kpc})^{2}],
\label{eq:mu}\end{equation}
where we have left the units of $\rm{mag\ arcsec}^{-2}$ implicit. Together, these detection criteria depend on the absolute magnitude, size, and radial distance of each mock satellite.

Figure \ref{fig:vis} demonstrates that our predicted satellite populations agree fairly well with the abundance and properties of classical and SDSS-identified systems; we discuss the apparent deficit of satellites in the inner regions as follows.Although Figure \ref{fig:vis} shows a particular realization of our model for a single MW-like host halo, we have checked that this mock satellite population is representative of our predictions for classical and SDSS systems. Our predicted satellite populations are similar to those inferred from semi-analytic models that account for satellite sizes, including \cite{Li09091291}, although we generally find fewer bright, compact systems than these works. Interestingly, the population of undiscovered mock satellites that do not pass the SDSS detection criteria depends sensitively on their size distribution. In particular, non-observations above the SDSS detection threshold in the $M_V$--$r_{\rm{\odot}}$ plane would be counted as detections in models that assume all dwarfs have sufficient surface brightness to be detected. This result suggests that size modeling will play an important role in interpreting current and future MW satellite observations. In addition, Figure \ref{fig:vis} shows that the number of systems that do not pass the SDSS detection thresholds depends strongly on $\mathcal{M}_{\rm{min}}$, which implies that the observed abundance and properties of MW satellites can be used to place upper limits on the masses of the subhalos that host faint systems (e.g., \citealt{Graus180803654,Jethwa161207834}).

Our fiducial model underpredicts the number of observed satellites at small radii ($r_{\rm{\odot}}\lesssim 50\ \rm{kpc}$) relative to SDSS observations. Decreasing the strength of baryonic subhalo disruption does not directly resolve this issue, since we predict too few surviving subhalos in this regime regardless of their disruption probabilities, which can be seen from Figure \ref{fig:vis}. However, our simulations are likely subject to spurious subhalo disruption and halo finder issues for subhalos in central regions, so it is unclear whether the difference among the predicted and observed radial satellite distributions reflects a physical shortcoming of our model or a numerical shortcoming of our simulations. We find that increasing the contribution from orphan satellites (i.e., increasing $\mathcal{O}$) alleviates the discrepancy, which hints at the latter explanation, although scaling our satellite radii inward (i.e., decreasing $\chi$) also reduces the tension. We revisit the effects of these parameters on the radial distribution in Section \ref{lowmass}.

\subsection{Quantitative Comparison}
\label{fit}

We now describe our procedure for fitting the model described in Section \ref{model} to observed satellite populations; we then specialize to the luminosity distribution of classical and SDSS-identified MW satellites. For each zoom-in simulation and each realization of our satellite model, we generate a predicted satellite population by performing a mock survey with an appropriate footprint and detection efficiency. We bin these satellites according to their physical properties $\mathbf{s}$ (e.g., absolute magnitude, heliocentric distance, and half-light radius) by discretizing the space of satellite properties into bins of volume $\mathcal{V}$; for example, choosing uniform bins for the parameter space plotted in Figure \ref{fig:vis} would correspond to setting $\mathcal{V} = \Delta M_V\Delta r_{\rm{\odot}} \Delta r_{1/2}$, where, for example, $\Delta M_V$ denotes the width of our absolute magnitude bins. To include the effects of observational incompleteness and satellite disruption, we count each satellite as $p_{\rm{detect}}\times(1-p_{\rm{disrupt}})$ --- rather than one --- observed system, where $p_{\rm{detect}}$ is the probability of detecting a given satellite (determined by its properties and the survey sensitivity) and $p_{\rm{disrupt}}$ is its disruption probability.

We assume that observed satellites and mock satellites populate the parameter space in question according to a multidimensional Poisson point process with a rate parameter $\lambda$ that is constant in each bin. Naively, $\lambda$ can be calculated by averaging the number of mock satellites in each bin:
\begin{equation}
\lambda_i(\boldsymbol{\theta}) = \frac{1}{\mathcal{V}}\langle \hat{n}_i(\boldsymbol{\theta})\rangle,\label{eq:poisson}
\end{equation}
where $\hat{n}_i(\boldsymbol{\theta})$ indicates the number of mock observed satellites in bin $i$, $\boldsymbol{\theta}$ denotes the set of model parameters, and the average can be taken over different zoom-in simulations and realizations of the satellite population in each simulation (including different luminosity and size model realizations, observer locations, and survey orientations). In this formulation, the likelihood of observing $N$ satellites $\mathbf{s}_1,\dots,\mathbf{s}_{N}$ in the survey corresponding to these mock observations is given by
\begin{equation}
P(\mathbf{s}_1,\dots,\mathbf{s}_{N}|\boldsymbol{\theta}) \approx \exp\Big[-\sum_{\text{bins}\ j} \lambda_j(\boldsymbol{\theta})\mathcal{V}_j\Big]\prod_{\text{bins}\ i}\frac{\lambda_i(\boldsymbol{\theta})^{n_{i}}}{n_{i}!},\label{eq:likelihood}
\end{equation}
where $i$ and $j$ index the bins and $n_{i}$ is the number of observed satellites in bin $i$. This expression is approximate because the integral that appears in the normalization factor for a Poisson point process is replaced by a sum over discrete bins.

However, the estimate of $\lambda$ obtained from Equation \ref{eq:poisson} is potentially noisy because we use a finite number of independent simulations. In addition, although the stochasticity in our predicted satellite populations is reduced because we restrict our analysis to host halos with two Magellanic Cloud analogs, it is necessary to impose a small lower bound on $\lambda$ to avoid realizations with zero likelihood if we adopt Equations \ref{eq:poisson}--\ref{eq:likelihood}. This choice of lower bound is necessarily arbitrary.

We therefore approach the problem in a different manner. Rather than calculating a single estimate of the rate parameter in each bin, $\lambda_i \left( \boldsymbol{\theta} \right)$, from the mock observed satellites, we marginalize over an unknown rate parameter in each bin. In particular, if we observe $n_i$ real satellites and $\hat{n}_{i,j}$ mock satellites in bin $i$, where $j=1,\dots ,\hat{N}$ runs over all simulations and model realizations, we have
\begin{align}
&P(n_{i}|\hat{n}_{i,1},\dots,\hat{n}_{i,\hat{N}}) = \int P(n_i|\lambda_i)P(\lambda_i|\hat{n}_{i,1},\dots,\hat{n}_{i,\hat{N}})\ \text{d}\lambda_i&\nonumber \\
&=\frac{1}{P(\hat{n}_{i,1},\dots,\hat{n}_{i,\hat{N}})}\int P(n_i|\lambda_i)P(\hat{n}_{i,1}|\lambda_i)\cdots P(\hat{n}_{i,\hat{N}}|\lambda_i)P(\lambda_i)\ \text{d}\lambda_i &\nonumber \\
&= \Big(\frac{\hat{N}+1}{\hat{N}}\Big)^{-(\hat{n}_{i,1} + \dots + \hat{n}_{i,\hat{N}} + 1)}\nonumber \\ &\times (\hat{N}+1)^{-n_i}\frac{(\hat{n}_{i,1} + \dots + \hat{n}_{i,\hat{N}} + n_i)!}{n_i!(\hat{n}_{i,1} + \dots + \hat{n}_{i,\hat{N}})!},&\label{eq:like}
\end{align}
where we have left the dependence on the model parameters $\boldsymbol{\theta}$ implicit, and we have assumed (i) a flat prior on $\lambda_i$ for $\lambda_i\geqslant 0$ and (ii) that $n_i$ and all $\hat{n}_{i,j}$ are drawn from the same Poisson distribution with rate parameter $\lambda_i$.

Because we produce non-integer numbers of mock satellites by counting each system as $p_{\rm{detect}}\times (1-p_{\rm{disrupt}})$ object, we replace the factorials in Equation \ref{eq:like} with the appropriate Gamma functions to obtain the final form of the likelihood. Our results are unaffected if we enforce integer counts by performing multiple mock observations of each predicted satellite population.

We note that Equation \ref{eq:like} is similar to the likelihood derived in \cite{Jethwa161207834}; however, these authors marginalize over $\lambda_i$, given a single predicted satellite population from a particular simulation and model realization, and then average the resulting probabilities, while our likelihood treats all simulations and model realizations simultaneously. In Appendix \ref{appendixa}, we demonstrate that our likelihood converges to the underlying Poisson distribution in the limit of many mock observations, while the likelihood used in \cite{Jethwa161207834} does not.

Finally, we use Bayes' theorem to compute the resulting posterior distribution over our free parameters:
\begin{equation}
P(\boldsymbol{\theta}|\mathbf{s}_1,\dots,\mathbf{s}_N) = \frac{P(\mathbf{s}_1,\dots,\mathbf{s}_N|\boldsymbol{\theta})P(\boldsymbol{\theta})}{P(\mathbf{s}_1,\dots,\mathbf{s}_N)},
\end{equation}
where $P(\boldsymbol{\theta})$ is our prior distribution, $P(\mathbf{s}_1,\dots,\mathbf{s}_N)$ is the Bayesian evidence, and
\begin{equation}
    P(\mathbf{s}_1,\dots,\mathbf{s}_N|\boldsymbol{\theta})=\prod_{\text{bins}\ i} P(n_{i}|\hat{n}_{i,1},\dots,\hat{n}_{i,\hat{N}})
\end{equation}
is the likelihood.

For our fit to classical and SDSS-identified satellites, we use the six MW-like host halos described previously. For each host, we generate five mock satellite populations by simultaneously drawing (i) satellite luminosities and sizes from our luminosity and size relations, (ii) random observer locations $8\ \rm{kpc}$ from the host halo center from the vertices of an octahedron, and (iii) random survey orientations. We assume that mock classical satellites ($M_V\leq -8.8\ \rm{mag}$) over the entire sky are always detected, and that mock satellites in the SDSS footprint that pass both of the detection criteria described in Section \ref{visualization} are always detected, and we fit our model to the luminosity distribution of the classical and SDSS-identified systems listed in Section \ref{visualization}. We choose uniform absolute magnitude bins, and we assume that $P(\boldsymbol{\theta})$ factorizes into a product of independent prior distributions; we list our choices for these priors in Table \ref{tab:prior}. We have checked that our results are not significantly affected by our choice of magnitude bins and priors.

Due to the limited constraining power of the classical-plus-SDSS luminosity distribution, we only vary the following parameters: $\alpha$, $\sigma_M$, $\mathcal{M}_{\rm{min}}$, and $\mathcal{B}$. We fix $\chi=1$, $\mathcal{A} = 0.02$ and $\gamma = -0.7$ (i.e., the fiducial \citealt{Jiang180407306} size relation), $\beta = 1$ (i.e., our fiducial satellite size reduction model), $\sigma_R = 0.01\ \rm{dex}$, and $\mathcal{O} = 1$ (i.e., our fiducial orphan model) for the fit presented here. Our choice of $\sigma_R$ is motivated by the fact that larger values of this scatter produce an overabundance of small observed mock satellites, since systems that scatter to small sizes at fixed distance and absolute magnitude are more likely to be observed. Thus, implementing an appreciable amount of size scatter likely requires modifying our assumptions that satellite size follows a lognormal distribution at fixed subhalo properties and that the scatter in our size relation is size-independent. 
In Appendix~\ref{appendix}, we test whether our radial scaling, size reduction, and orphan models are preferred by the classical-plus-SDSS luminosity distribution by computing Bayes factors for fits with $\chi=1$, $\beta = 1$, and $\mathcal{O} = 1$ versus fits with $\chi=0.8$, $\beta = 0$, and $\mathcal{O} = 0$. We find very weak evidence in favor of $\chi=0.8$, $\beta=1$, and $\mathcal{O}=1$, implying that our fit to classical-plus-SDSS satellites is not sensitive to these effects and justifying our choice to fix these parameters.

\begin{center}
\begin{table*}[t]
\centering
\begin{tabular}{l|l|l}
\hline
Free Parameter & Prior Distribution & Bounds/Shape of Prior Set By \\ \hline \hline
Faint-End Slope ($\alpha$)
& $\arctan\alpha\sim\text{unif}(-1.1,-0.9)$ 
& Uninformative prior for $-2\lesssim\alpha\lesssim -1.25$.\\
\hline
Luminosity Scatter ($\sigma_M$)
& $\sigma_M\sim\text{unif}(0.2\ \rm{dex},1.0\ \rm{dex})$ 
& $\sigma_M\approx 0.2\ \rm{dex}$ for higher-mass halos.\\
\hline
Galaxy Formation Threshold ($\mathcal{M}_{\rm{min}}$)
& $\log(\mathcal{M}_{\rm{min}}/\msun)\sim\text{unif}(7.5,10)$ 
& Conservative upper bound based on \cite{Jethwa161207834}.\\
\hline
Baryonic Subhalo Disruption ($\mathcal{B}$)
& $\mathcal{B}\sim$ Lognormal$(\mu=1,\sigma=0.5)$ 
& $\mathcal{B} = 1$ corresponds to hydrodynamic results.\\
\hline
\end{tabular}
\caption{Prior distributions for the parameters varied in our fit to the luminosity distribution of classical and SDSS-identified MW satellites.}
\label{tab:prior}
\end{table*}
\end{center}

To sample the posterior distribution for our fit to the classical-plus-SDSS luminosity distribution, we run $5\times 10^{4}$ iterations of the Markov Chain Monte Carlo (MCMC) sampler \texttt{emcee} \citep{emcee} using $20$ walkers, and we discard the first $5000$ burn-in steps. We have verified that our results are stable to changes in the number of walkers and that we have sampled a reasonable number of autocorrelation lengths for each chain.


\section{Results and Discussion}
\label{Results}

\begin{figure*}
\centering
\includegraphics[scale=0.45]{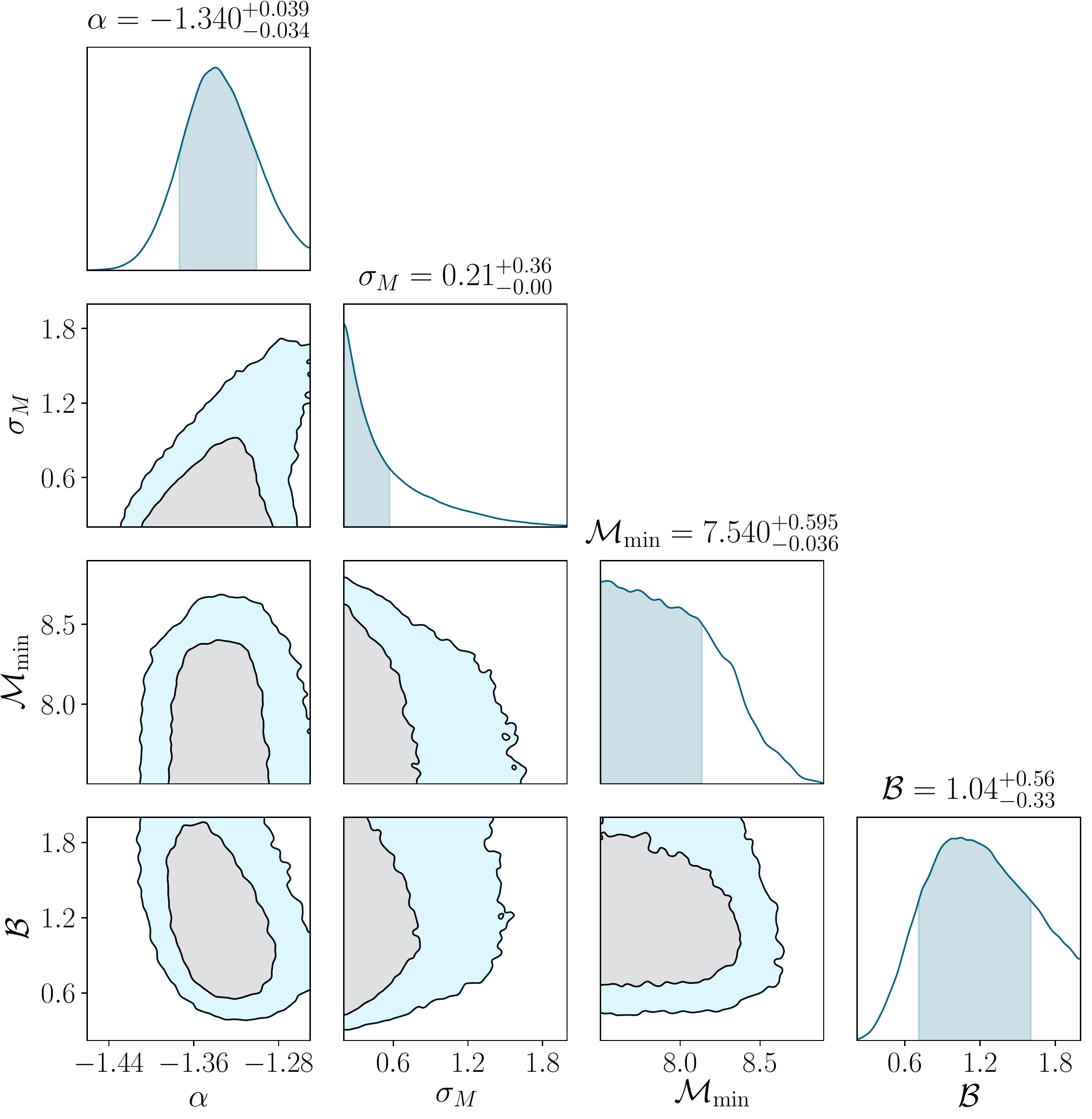}
\caption{Posterior distribution from our fit to the luminosity distribution of classical and SDSS-identified satellites. Dark (light) shaded contours show $68\%$ ($95\%$) confidence intervals, and shaded areas in the marginal distributions show $68\%$ confidence intervals. Note that $\sigma_M$ is reported in $\text{dex}$ and $\mathcal{M}_{\rm{min}}$ is reported as $\log(\mathcal{M}_{\rm{min}}/\msun)$ in this plot.}
\label{fig:corner}
\end{figure*}

The posterior distribution from our MCMC run is shown in Figure \ref{fig:corner}; we now summarize our main results.

\subsection{Derived Constraints}

\begin{enumerate}
\item Our fit favors a faint-end luminosity function slope of $\alpha=-1.34^{+0.04}_{-0.03}$ ($68\%$ confidence interval), which is shallower than most constraints from \cite{Tollerud08064381} and fairly consistent with the global constraint from GAMA ($\alpha=-1.26 \pm 0.07$; \citealt{Loveday150501003}).\footnote{Note that \cite{Loveday150501003} constrain the Schechter function faint-end slope while we measure the power-law luminosity function slope itself.} The GAMA constraint is derived from the luminosity function of galaxies with $-24\ \text{mag}\lesssim M_r\lesssim -13\ \rm{mag}$, while we constrain the power-law slope of the luminosity function for systems dimmer than $M_r = -13\ \rm{mag}$ by fitting to satellites with $-18\ \text{mag}\lesssim M_V\lesssim -1\ \rm{mag}$.
\item Our fit is consistent with the luminosity scatter inferred for higher-mass galaxies ($\sigma_M \approx 0.2\ \rm{dex}$), but it allows for significantly larger values, with $\sigma_M = 0.21^{+0.36}_{-0.00}\ \rm{dex}$ ($68\%$ confidence interval).
\item Our fit strongly favors a galaxy formation threshold of $\mathcal{M}_{\rm{min}} < 5\times 10^{8}\ \msun$, with $\log(\mathcal{M}_{\rm{min}}/\msun)=7.54^{+0.60}_{-0.04}$ ($68\%$ confidence interval); our results are consistent with the upper bound of $2.4 \times 10^{8}\ \msun$ from~\cite{Jethwa161207834}. Decreasing $\mathcal{M}_{\rm{min}}$ below $10^8\ \msun$ rarely results in additional mock classical or SDSS observations, so the marginal likelihood for $\mathcal{M}_{\rm{min}}$ is roughly flat in this regime.
\item Our fit is consistent with $\mathcal{B} = 1$: we find $\mathcal{B}=1.04^{+0.56}_{-0.33}$ ($68\%$ confidence interval), which implies that our fiducial baryonic disruption model is compatible with the observed classical-plus-SDSS luminosity distribution.
\end{enumerate}

\begin{figure}
\centering
\includegraphics[scale=0.35]{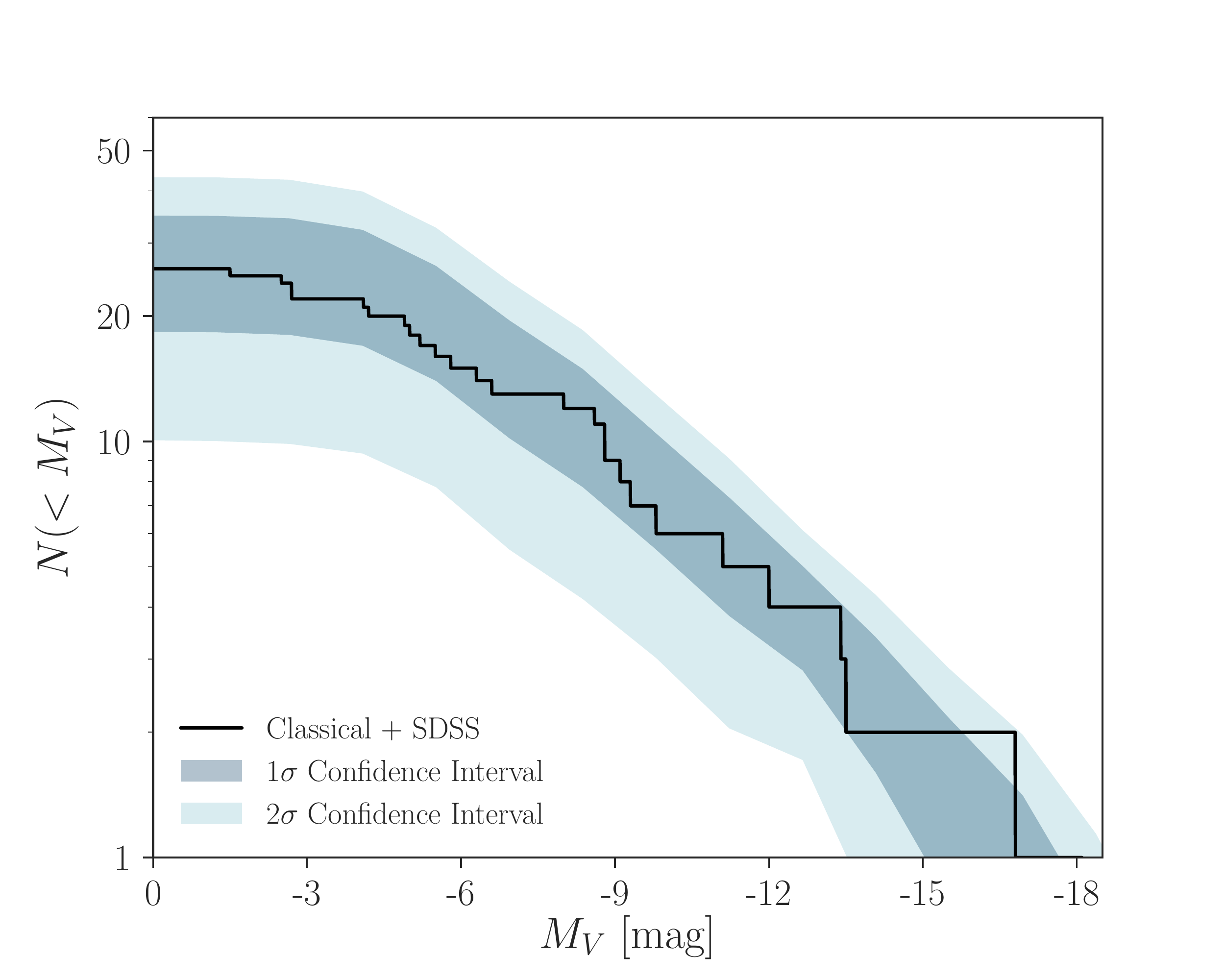}
\caption{Our prediction for the observed luminosity function of classical and SDSS satellites inferred from our fit to the absolute magnitude distribution of these systems. Dark (light) shaded areas show $68\%$ ($95\%$) confidence intervals.}
\label{fig:LF}
\end{figure}

To test whether our model provides an adequate fit to the data, we draw samples from the posterior distribution and plot the resulting $68\%$ and $95\%$ confidence intervals for the luminosity distribution, radial distribution, and size distribution of classical and SDSS satellites in Figures \ref{fig:LF} and \ref{fig:radialdist}. Our predictions are largely consistent with both the observed luminosity function and the observed radial and size distributions of these systems, despite the fact that we have only fit to their luminosities. As noted previously, our model slightly underpredicts the observed population of satellites close to the center of the MW ($r_{\rm{\odot}}\lesssim 50\ \rm{kpc}$). However, since we have only fit to an observed luminosity distribution using fixed radial scaling and orphan prescriptions, this discrepancy might not persist for a joint fit to observed satellite luminosities, radii, and sizes that varies $\chi$ and $\mathcal{O}$. The dashed red lines in Figure \ref{fig:radialdist} illustrate that decreasing $\chi$ reduces the tension among the predicted and observed inner radial distributions.

\subsection{Predictions for Future Surveys}

\begin{figure*}
\centering
\includegraphics[scale=0.35]{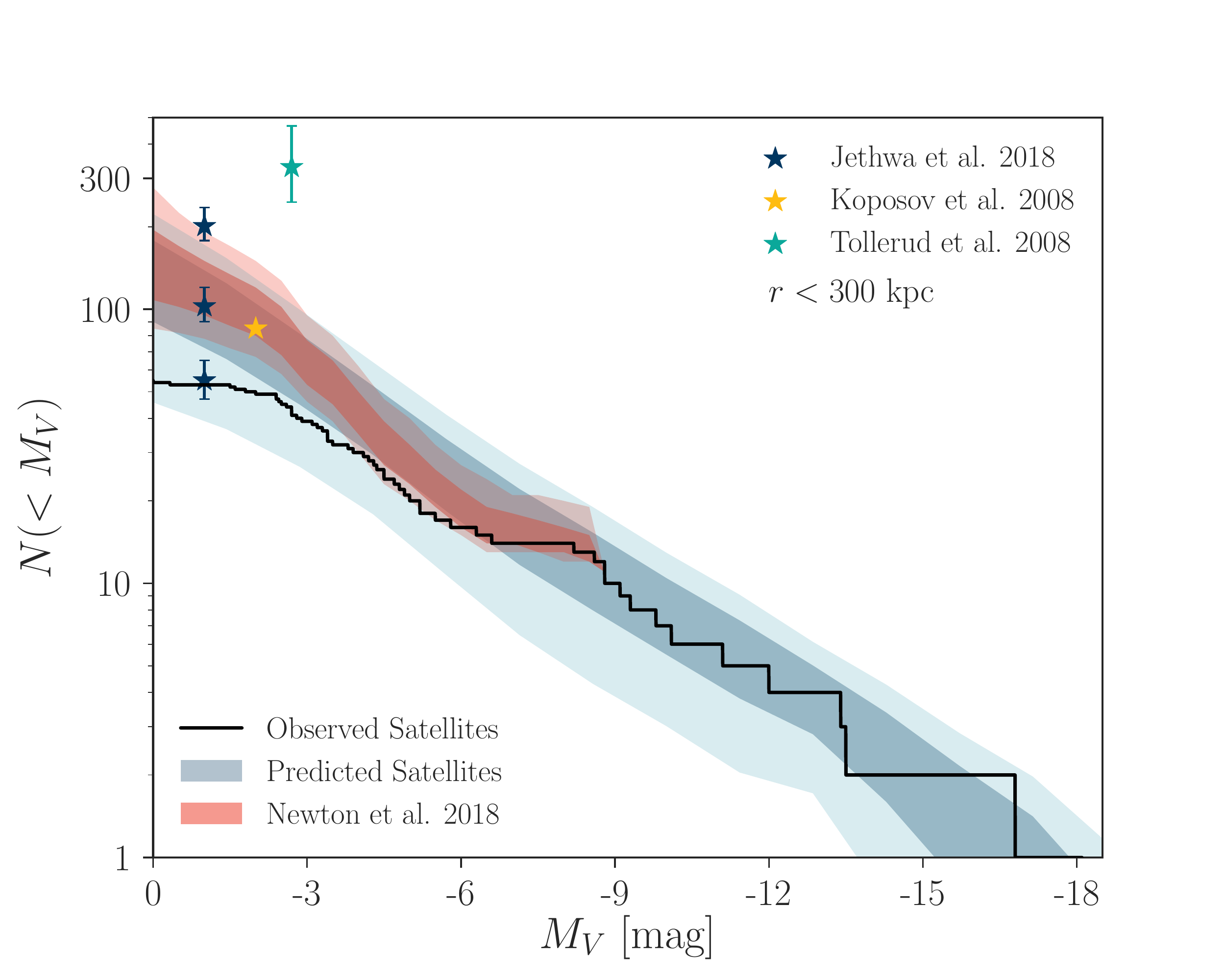}
\includegraphics[scale=0.35]{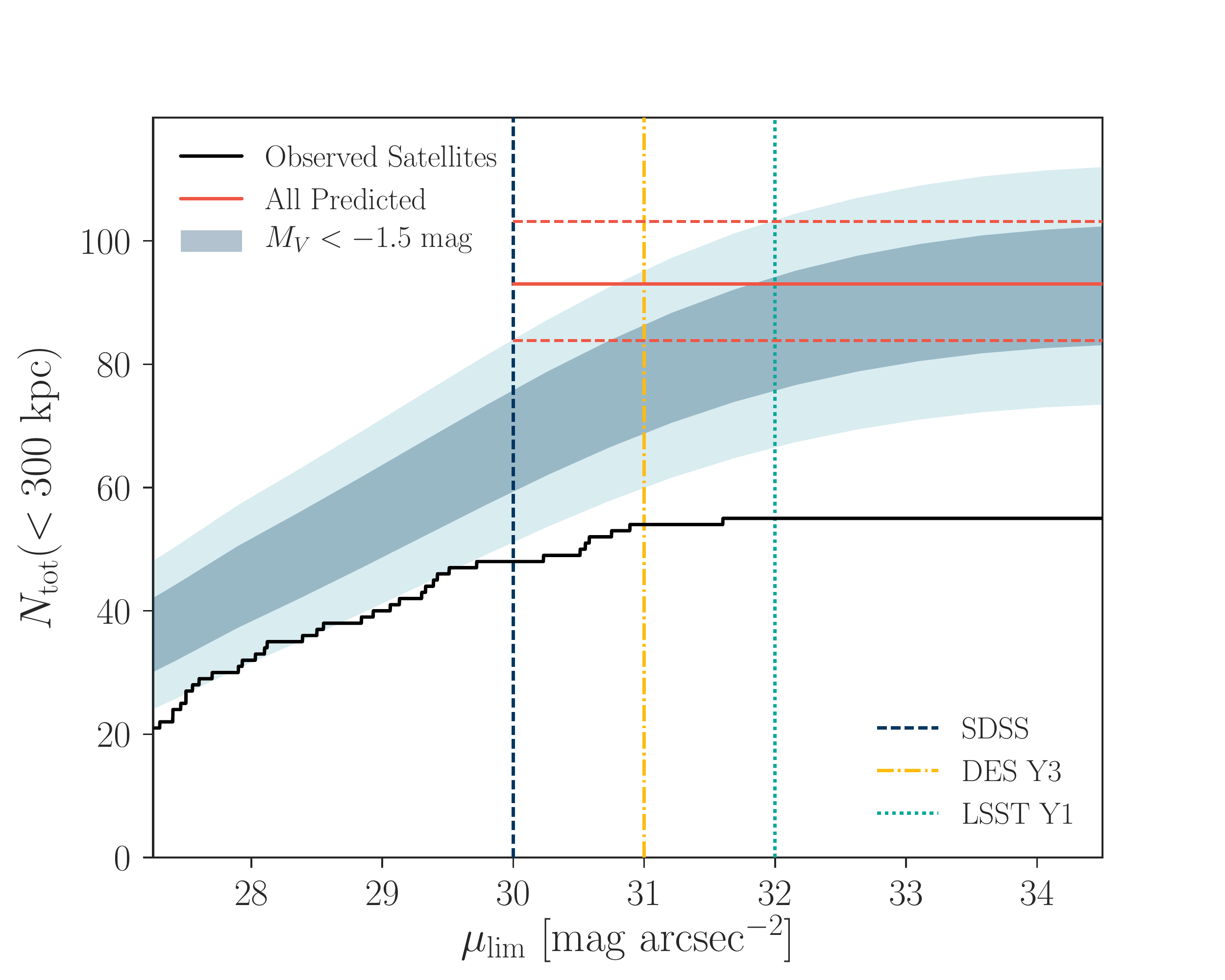}
\caption{Left panel: the total number of satellites within $300\ \rm{kpc}$ of the MW inferred from our fit to the classical-plus-SDSS luminosity distribution (blue) compared to results from previous studies and to all observed MW satellites including candidate systems (black). Right panel: the total number of satellites with $M_V<-1.5\ \rm{mag}$ within $300\ \rm{kpc}$ of the MW as a function of limiting observable surface brightness. The red line shows our prediction for the total number of satellites with $M_V<-1.5\ \rm{mag}$ independent of surface brightness; dashed red lines show $68\%$ Poisson confidence intervals. In both panels, dark (light) shaded areas indicate $68\%$ ($95\%$) confidence intervals.}
\label{fig:lf_tot}
\end{figure*}

Given the sky coverage and detection efficiency of future MW satellite searches, we can use our model to predict the abundance and properties of the satellites that we expect to be discovered. To place our results in context, we first study the total number of satellites within $300\ \rm{kpc}$ of the MW --- independent of surface brightness --- inferred from our fit to the classical-plus-SDSS luminosity distribution. The left-hand panel of Figure \ref{fig:lf_tot} compares our prediction for the total number of MW satellites as a function of absolute magnitude to the results in \cite{Tollerud08064381}, \cite{Koposov07062687}, and \cite{Newton170804247}, and to the predictions derived in \cite{Jethwa161207834} based on three different stellar mass--halo mass relations. We predict $95 \pm 29$ ($134 \pm 44$) total satellites with $M_V < -1.5\ \rm{mag}$ ($M_V < 0\ \rm{mag}$) within $300\ \rm{kpc}$ of the MW at the $68\%$ confidence level. Our estimate for the total number of MW satellites is consistent with but more conservative than most previous results, likely due to the fact that our subhalo disruption model captures both the effects of a central disk and additional baryonic physics. We refer the reader to \cite{Newton170804247} for a discussion of the discrepancy between the \cite{Tollerud08064381} prediction and other estimates. Finally, we note that the high-$\mathcal{M}_{\rm{min}}$ tail of our posterior results in a small number of realizations with fewer total satellites than currently observed (including both spectroscopically confirmed systems and candidate satellites); we choose not to incorporate this constraint in our fit to restrict our analysis to classical and SDSS data alone.

Next, we make predictions for satellite searches with improved surface brightness limits by calculating the total number of MW satellites as a function of limiting observable surface brightness. In particular, the right-hand panel of Figure~\ref{fig:lf_tot} shows the total number of MW satellites within $300\ \rm{kpc}$ inferred from our fit to the classical-plus-SDSS luminosity distribution, assuming that satellites over the entire sky are observed to a limiting surface brightness $\mu_{\rm{lim}}$. We predict that $83\pm 26$ ($92\pm 29$) satellites with $M_V < -1.5\ \rm{mag}$ would be observed if the entire sky were covered to a limiting surface brightness of~$32\ (34)\ \rm{mag\ arcsec}^{-2}$. Similarly, we predict that $\sim 95\%$ of all MW satellites with $M_V < -1.5\ \rm{mag}$ would be observed if satellites down to $33\ \rm{mag\ arcsec}^{-2}$ were observed over the entire sky. We emphasize that these estimates depend on the details of our size model and surface brightness calculations.

To connect these predictions to ongoing and future surveys, we indicate approximate surface brightness detection thresholds for SDSS, DES, and LSST in Figure \ref{fig:lf_tot}. We estimate the sensitivity of LSST satellite searches by comparing the $5\sigma$ limiting point-source magnitudes of a recent HSC-SSP satellite search \citep{Homma170405977} to those expected for LSST. LSST will likely achieve comparable sensitivity to this HSC-SSP survey in its first year of operation \citep{Ivezic08052366}; Equation \ref{eq:mu} implies that the two satellite candidates recently detected in HSC-SSP data have $\mu_V = 31.6\ \rm{mag\ arcsec}^{-2}$ (Virgo I) and $\mu_V = 30.9\ \rm{mag\ arcsec}^{-2}$ (Cetus III), so we adopt an approximate surface brightness threshold of $32\ \rm{mag\ arcsec}^{-2}$ for LSST Y1 (K.\ Bechtol 2019, private communication). We note that satellite searches with surveys such as HSC-SSP and LSST face the unique challenge of distinguishing faint stars associated with dwarf satellites in the Galactic halo from distant unresolved galaxies (e.g., \citealt{Willman09074758}), adding to the uncertainty in our limiting surface brightness estimate.\\

\subsection{Implications for the Connection between Low-mass Subhalos and Faint Satellites}
\label{lowmass}

\begin{figure}
\centering
\includegraphics[scale=0.4]{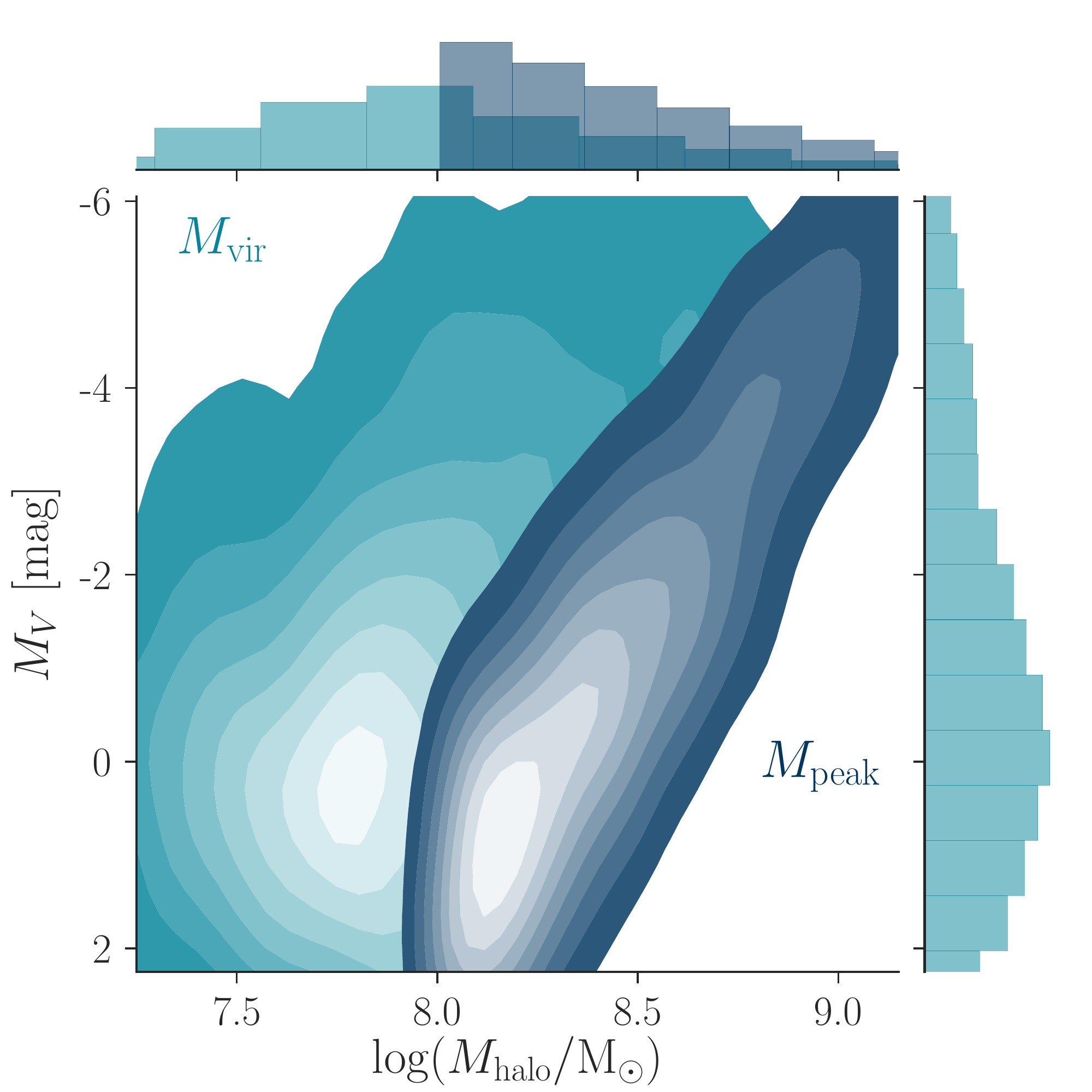}
\caption{Joint distribution of subhalo mass and satellite absolute magnitude predicted by our best-fit model with $\mathcal{M}_{\rm{min}}=10^{8}\ \msun$. The dark (light) contour corresponds to peak (present-day) subhalo virial mass, and the contours indicate satellite number density.}
\label{fig:Mhalo}
\end{figure}

The properties of the subhalos that host faint MW satellites can be used to constrain DM models that produce a cutoff in the subhalo mass function (e.g., \citealt{Maccio09102460,Kennedy13107739,Jethwa161207834}), along with the impact of reionization on galaxy formation (e.g., \citealt{Munoz09054744,Graus180803654}). Although we have not explicitly imposed a relationship between the luminosity of our mock satellites and the present-day mass of their subhalos, our model can be used to predict the joint distribution of subhalo mass and satellite luminosity, which we illustrate in Figure \ref{fig:Mhalo} by populating our MW-like host halos using the best-fit model derived previously with fixed $\mathcal{M}_{\rm{min}}=10^{8}\ \msun$. 
The tight correspondence between peak circular velocity and absolute magnitude enforced by our $M_V$--$V_{\rm{peak}}$ relation is broadened by the mass--concentration relation (which relates $V_{\rm{peak}}$ to $M_{\rm{peak}}$) and further broadened by tidal stripping (which relates $M_{\rm{peak}}$ to $M_{\rm{vir}}$).

\begin{figure*}
\centering
\includegraphics[scale=0.35]{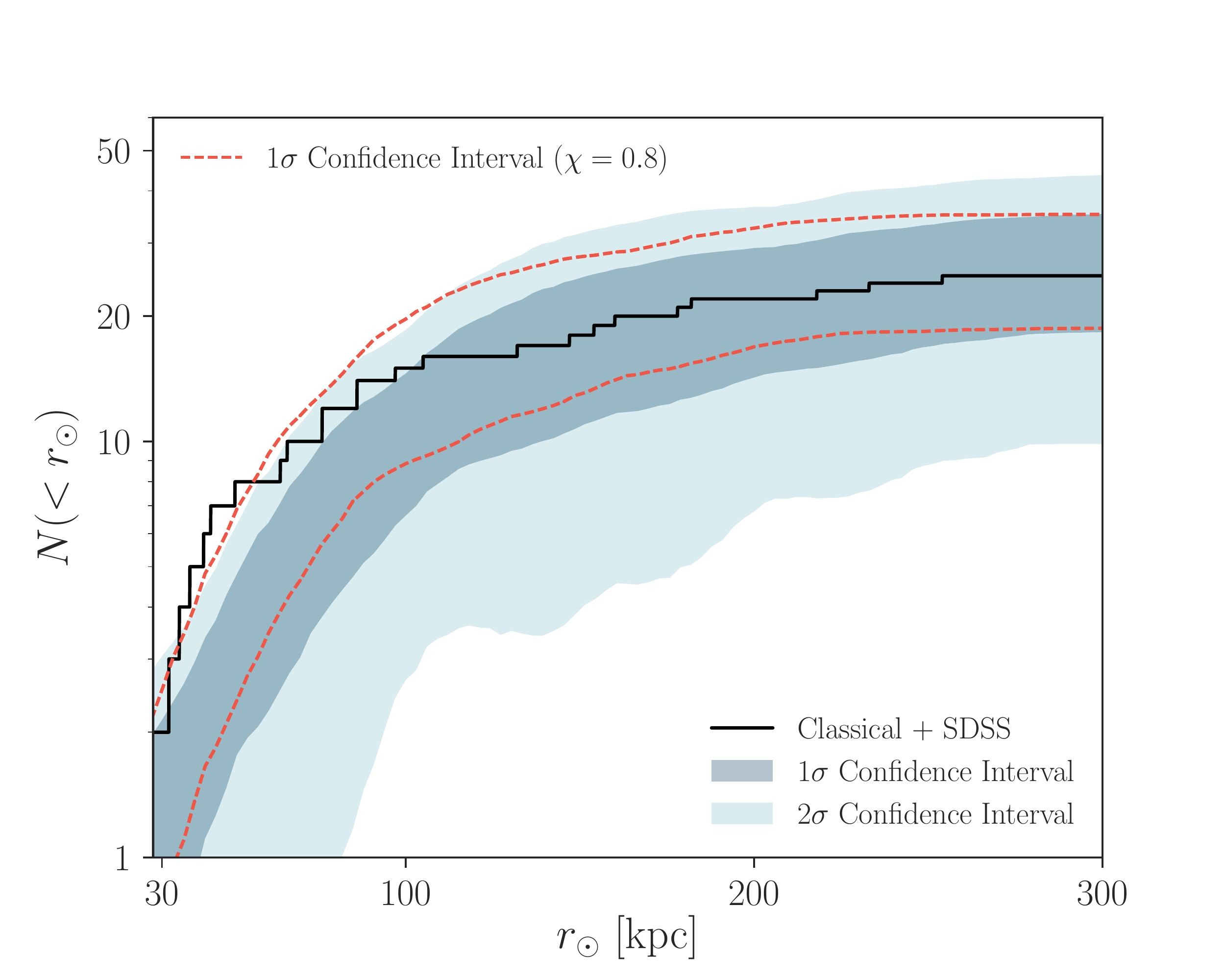}
\includegraphics[scale=0.35]{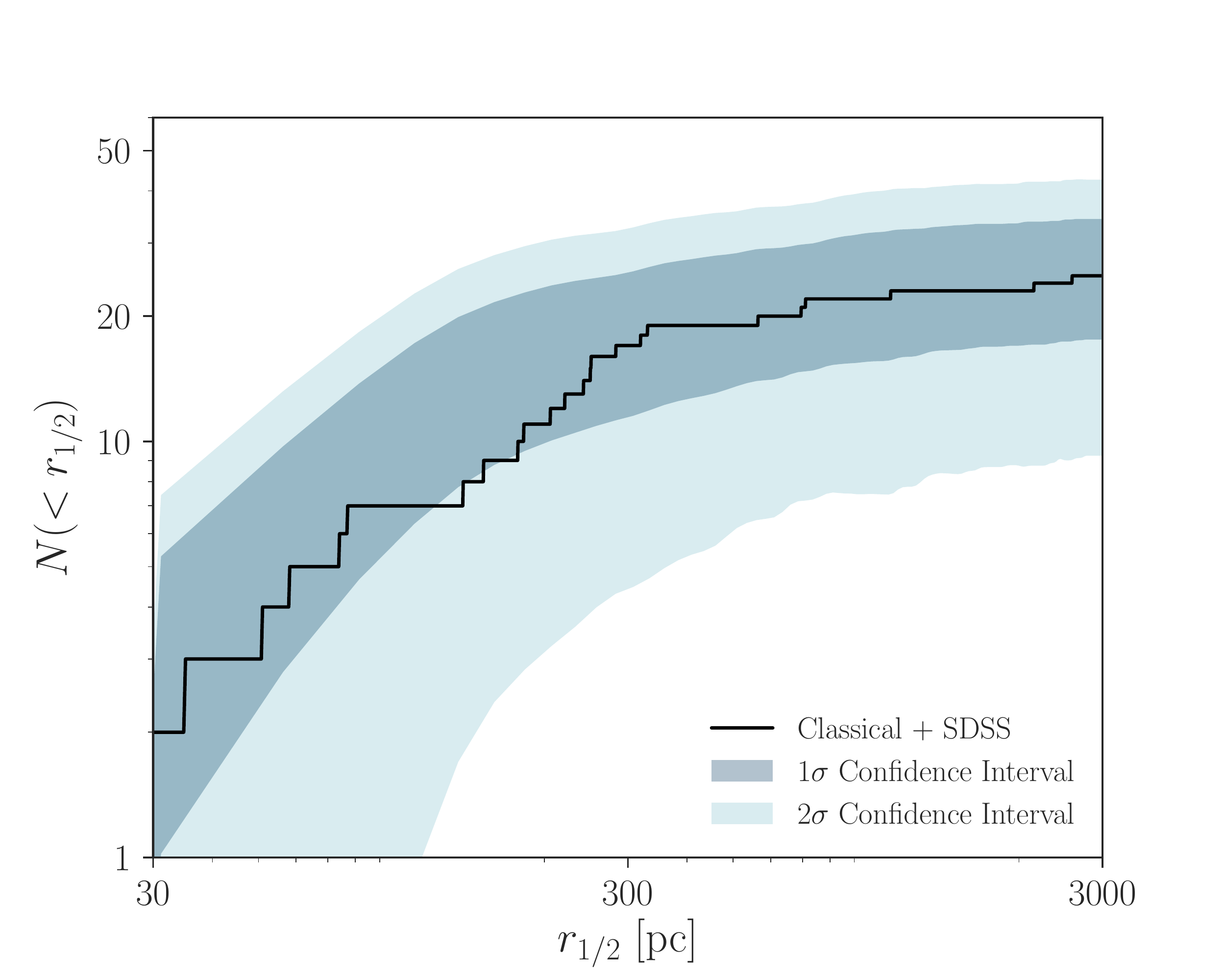}
\caption{Left panel: the radial distribution of classical and SDSS satellites inferred from our fit to the observed luminosity distribution of these systems. Dashed red lines show $68\%$ confidence intervals for a fit with satellite radii scaled inward relative to subhalo radii by a factor of $0.8$. Right panel: the corresponding size distribution, calculated by setting satellite sizes based on subhalo properties at accretion ($\beta = 0$) with a constant lognormal scatter of $\sigma_R = 0.01\ \rm{dex}$. In both panels, dark (light) shaded areas show $68\%$ ($95\%$) confidence intervals.}
\label{fig:radialdist}
\end{figure*}

By drawing from our full posterior distribution (i.e., allowing $\mathcal{M}_{\rm{min}}$ to vary) and repeatedly populating our six MW-like host halos, we find that satellites with $-6\ \text{mag}\lesssim M_V \lesssim 0\ \text{mag}$ typically reside in subhalos with present-day virial masses $2\times 10^{7}\ \msun\lesssim M_{\rm{vir}} \lesssim 2\times 10^{8}\ \msun$ and peak virial masses $10^{8}\ \msun\lesssim M_{\rm{peak}} \lesssim 5\times 10^{8}\ \msun$ ($68\%$ confidence intervals). We caution that baryonic effects can systematically reduce subhalo masses \citep{Okamoto08060378,Munshi12091389,Sawala14043724}, so these results should be interpreted as subhalo masses \emph{in DMO simulations}; an additional layer of modeling calibrated on hydrodynamic simulations would be necessary to modify these values to account for the presence of baryons. The quoted lower bounds depend mildly on the low-mass cutoff in our prior for $\mathcal{M}_{\rm{min}}$, since mock satellites in low-mass subhalos can scatter into the absolute magnitude range of interest; however, we caution that our adopted low-mass cutoff is near the resolution limit of our simulations. Our upper bounds are consistent with the results in \cite{Jethwa161207834}, who showed that mass thresholds in this range can be used to place competitive constraints on warm dark matter models.

We find that the faint satellites considered previously typically inhabit subhalos with $14\ \text{km\ s}^{-1}\lesssim V_{\rm{peak}}\lesssim 22\ \rm{km\ s}^{-1}$ ($68\%$ confidence interval). This lower bound on $V_{\rm{peak}}$ can potentially decrease for a joint fit to the luminosities and radial distances of observed satellites; for example, \cite{Graus180803654} showed that it might be necessary to populate subhalos down to $V_{\rm{peak}}\approx 10\ \rm{km\ s}^{-1}$ to match the observed radial distribution of MW satellites. However, these bounds are also dependent on our radial scaling and orphan satellite models, since decreasing $\chi$ or increasing $\mathcal{O}$ raises the predicted abundance of mock satellites, particularly in the inner regions ($r\lesssim 50\ \rm{kpc}$). For example, by drawing from our posterior and repeatedly populating our MW-like host halos with fixed $\mathcal{O}=1$, we predict $9\pm 3$ satellites with $r<50\ \rm{kpc}$ and $M_V< 0\ \rm{mag}$ if satellite radii are set equal to subhalo radii ($\chi=1$), and $14\pm 5$ ($37\pm 12$) such satellites if $\chi=0.8$ ($\chi=0.5$). On the other hand, for fixed $\chi=1$, we predict $5\pm 2$ satellites with $r<50\ \rm{kpc}$ and $M_V< 0\ \rm{mag}$ if no orphans are included ($\mathcal{O}=0$), and $13\pm 4$ such systems if all orphans are included with zero disruption probability ($\mathcal{O}\gg 1$; 68\% confidence intervals). Our radial scaling and orphan parameters are therefore somewhat degenerate, but the fact that $18$ satellites with $r<50\ \rm{kpc}$ have been observed (including candidate satellites) implies that it is difficult to reconcile models with fixed $\chi=1$ and $\mathcal{O}=0$ with the observed radial distribution.

Thus, in the context of our model, it is unclear whether the observed radial distribution requires a lower galaxy formation threshold than expected in standard reionization quenching scenarios. Future work that addresses a larger population of faint MW satellites 
will shed light on this issue; for example, including the ultra-faint dwarfs discovered by DES \citep{Bechtol150302584,Drlica-Wagner150803622} in a joint fit to satellite luminosities and radii can potentially decrease our upper bound on $\mathcal{M}_{\rm{min}}$ and break the degeneracies necessary to constrain our radial scaling and orphan models. Such a study will require a careful treatment of LMC satellites, which is beyond the scope of this paper.

\subsection{Caveats and Future Work}
\label{caveats}

There are several caveats and possible extensions to the model presented in this paper. Most notably, the fact that we underpredict the number of satellites in the inner regions can potentially bias our parameter constraints and predictions for future surveys. However, we find that refitting the classical-plus-SDSS luminosity distribution with $\chi=0.8$ to alleviate this tension (see Figure \ref{fig:radialdist}) does not significantly affect our results.

Although we have focused on modeling the observed absolute magnitudes, radial distances, and physical sizes of MW satellites, stellar velocity dispersion measurements provide an additional constraint on the subhalos that host these systems. The $V_{\rm{max}}$ distribution of the subhalos that host our mock classical satellites extends to significantly higher values than those inferred from stellar velocity dispersion measurements using the \cite{Wolf09082995} $V_{\rm{max}}$ estimator; thus, our model suffers from the canonical ``too big to fail'' problem \citep{Boylan-Kolchin11030007,Boylan-Kolchin11112048}, despite the fact that we include subhalo disruption due to baryonic effects. This shortcoming can likely be mitigated by incorporating the effects of stellar feedback on the inner density profiles of subhalos, along with a more sophisticated conversion between $V_{\rm{max}}$ values and observed stellar velocity dispersions (\citealt{Zolotov12070007,Maccio160701028,Brooks170107835,Campbell160304443,Verbeke170303810}; see \citealt{Jiang150802715} and \citealt{Bullock170704256} for comprehensive discussions of the ``too big to fail" problem). 

By construction, our approach relies on zoom-in simulations; however, given an analytic method for generating subhalo $V_{\rm{peak}}$ functions, radial distributions, and size distributions, along with parameterizations of tidal stripping, subhalo disruption due to baryonic effects, and the contribution of orphan satellites, mock satellite populations could potentially be generated without using a particular set of simulations (see \citealt{Han150902175} for work along these lines). Such an approach would effectively combine our empirical framework with semi-analytic models to increase the level of detail at which we model satellite properties.

Exploring how our results depend on host halo mass would require either a different set of simulations or an analytic model, and characterizing this dependence might be particularly important in light of recent results that favor a relatively massive MW halo (\citealt{Monari180704565,Posti180501408,Simon180410230,Watkins180411348}; however, see \citealt{Callingham180810456}). In addition, although we have focused on satellite luminosities, radii, and sizes, comparing the orbital properties of observed and simulated systems will likely be fruitful in the era of precision astrometric measurements.


\section{Summary}
\label{Discussion}

We have described a flexible model for populating subhalos in DMO zoom-in simulations of MW-mass hosts with satellite galaxies. We demonstrated that this model produces reasonable satellite populations in the regime of classical and SDSS-identified MW satellites, and we presented an improved method for fitting the model to observed satellite populations. Our fit to the classical-plus-SDSS luminosity distribution produces satellite populations that are qualitatively and quantitatively consistent with the luminosity function, radial distribution, and size distribution of observed systems, modulo modest tension in the inner radial distribution.

We briefly summarize the key aspects of our approach and highlight several open questions:
\begin{enumerate}[wide, labelwidth=!, labelindent=0pt, label=({\roman*}), itemsep=0pt]
\item \emph{Host halo properties:} We fix host halo properties based on our zoom-in simulation suite. Our host halos lie in the mass range $10^{12.1\pm 0.03}\ \msun$ and have a variety of accretion histories and secondary properties. For our fit to observed MW satellites, we select host halos with two Magellanic Cloud-like systems. How do our predicted satellite populations vary as a function of host halo mass?
\item \emph{Satellite luminosities:} We assign satellite luminosities to DMO subhalos by abundance matching to the GAMA survey down to $M_r = -13\ \rm{mag}$ and extrapolating this relation to fainter systems assuming a power-law luminosity function. We set the threshold for galaxy formation due to reionization using a cut on peak subhalo virial mass. Are constraints on the subhalo--satellite connection derived from MW satellites consistent with those from M31, the Local Volume, and SAGA hosts? What galaxy formation threshold is consistent with observations of ultra-faint satellites discovered since SDSS? What is the lowest subhalo mass that future observations of MW satellites can probe?
\item \emph{Satellite locations:} We model the locations of our mock satellites by scaling the radial distances of subhalos in our zoom-in simulations and projecting these systems onto the sky. Is the seemingly centrally concentrated radial distribution of MW satellites rare, or is it an artifact of misestimated observational incompleteness?
\item \emph{Satellite sizes:} We assign satellite sizes to DMO subhalos using the size relation from \cite{Jiang180407306}. Is there evidence (e.g., from hydrodynamic simulations) that this size relation holds for ultra-faint dwarf galaxies?
\item \emph{Baryonic effects:} We model subhalo disruption due to baryonic physics, such as the tidal influence of a galactic disk, using a model calibrated on hydrodynamic simulations. Were MW subhalos and satellites tidally disrupted in a manner that is consistent with hydrodynamic results?
\item \emph{Orphan satellites:} We include orphan satellites by tracking the orbits and modeling the tidal stripping of disrupted subhalos in our simulations. If subhalo disruption is a numerical artifact (e.g., \citealt{VandenBosch180105427,VandenBosch171105276}), such that a significant population of disrupted subhalos should host observable satellite galaxies, can it be accounted for by modeling orphans? What is the relationship between subhalo disruption and satellite disruption?
\end{enumerate}


\acknowledgments

Our code and processed subhalo catalogs are available at \href{https://github.com/eonadler/subhalo_satellite_connection}{github.com/eonadler/subhalo\_satellite\_connection}; please contact the authors with additional data requests. We thank Keith Bechtol, Alex Drlica-Wagner, Marla Geha, Ari Maller, and Andrew Wetzel for useful discussions and for comments on the manuscript. 
This research received support from the National Science Foundation (NSF) under grant no.\ NSF AST-1517422, grant no.\ NSF PHY11-25915 through the Kavli Institute for
Theoretical Physics program ``The Small-Scale Structure of Cold(?)\ Dark Matter,''
and grant no.\ NSF DGE-1656518 through the NSF Graduate Research Fellowship received by E.O.N. 
Y.-Y.M.\ is supported by the Samuel P.\ Langley PITT PACC Postdoctoral Fellowship.
This research made use of computational resources at SLAC National Accelerator Laboratory, a U.S.\ Department of Energy Office; the authors thank the support of the SLAC computational team.  
This research made extensive use of \https{arXiv.org} and NASA's Astrophysics Data System for bibliographic information.

\software{
ChainConsumer \citep{ChainConsumer},
emcee \citep{emcee},
IPython \citep{ipython},
Jupyter (\http{jupyter.org}),
Matplotlib \citep{matplotlib},
NumPy \citep{numpy},
SciPy \citep{scipy}, 
Seaborn (\https{seaborn.pydata.org}).
}

\bibliographystyle{yahapj}
\bibliography{references,software}


\appendix

\section{Convergence of Improved Likelihood to the Poisson Distribution}
\label{appendixa}

The main assumption underlying the derivation of our likelihood (Equation \ref{eq:like}) is that the number of observed satellites and mock satellites in a given magnitude bin are drawn from the same Poisson distribution. Thus, in the limit of many mock observations, any likelihood that marginalizes over an unknown rate parameter should converge to the underlying Poisson distribution. This holds for the likelihood used in our analysis: in particular, if $\hat{n}_{i,1},\dots,\hat{n}_{i,\hat{N}}\sim \text{Poisson}(\lambda_i)$, then
\begin{equation}
\lim_{\hat{N}\rightarrow\infty}P(n_{i}|\hat{n}_{i,1},\dots,\hat{n}_{i,\hat{N}}) = \frac{e^{-\lambda_i}\lambda_i^{n_i}}{n_i !} = \text{Poisson}(\lambda_i),
\end{equation}
where the scaling with $n_i$ follows from Equation \ref{eq:like} using $\hat{n}_{i,1}+\dots + \hat{n}_{i,\hat{N}} = \hat{N}\lambda_i$ for large $\hat{N}$, and the prefactor follows from the fact that $P(n_{i}|\hat{n}_{i,1},\dots \hat{n}_{i,\hat{N}})$ is a normalized distribution or by taking the $\hat{N}\rightarrow \infty$ limit of Equation \ref{eq:like}.

\begin{figure}[h]
\centering
\includegraphics[scale=0.55]{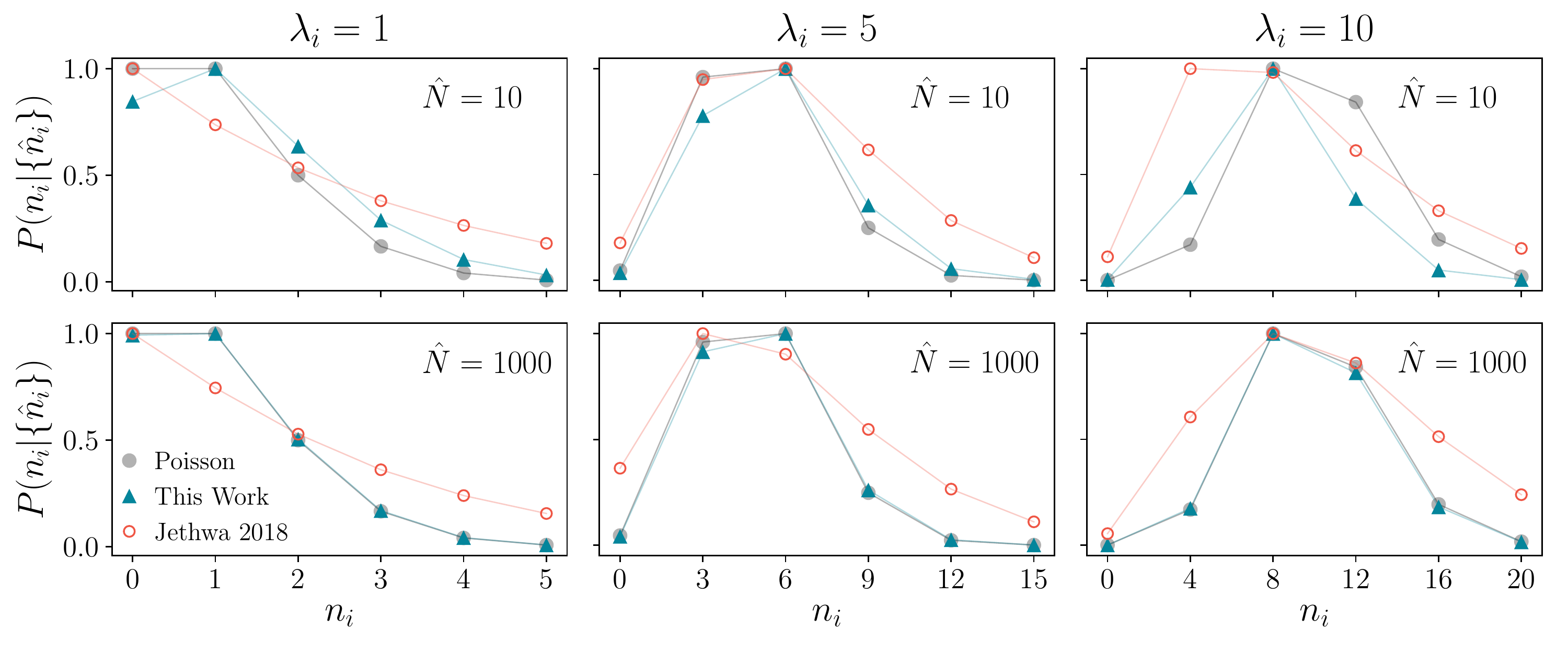}
\caption{Likelihood of observing $n_i$ counts given $\hat{N}$ mock observations $\hat{n}_{i,1},\dots ,\hat{n}_{i,\hat{N}}$ assuming that $n_i$ and all $\hat{n}_i$ are drawn from a Poisson distribution with rate parameter $\lambda_i$. We show results for $\hat{N}=10$ (top row) and $\hat{N}=1000$ (bottom row) mock observations given true rate parameters $\lambda_i=1$ (left column), $\lambda_i=5$ (middle column), and $\lambda_i=10$ (right column), computed using (i) the Poisson likelihood given the true rate parameter (gray points); (ii) the likelihood used in our analysis (Equation \ref{eq:like}), which marginalizes over $\lambda_i$ given all $\hat{n}_i$ simultaneously (blue triangles); and (iii) the likelihood used in \cite{Jethwa161207834}, which averages the likelihoods obtained from multiple mock observations (red circles). Our likelihood converges to the underlying Poisson distribution in the limit of many mock observations, while the averaged version does not. Note that we rescale each version of $P(n_i|\{\hat{n}_i\})$ by its maximum value in every panel.}
\label{fig:likelihoodplot}
\end{figure}

We illustrate the fact that our likelihood satisfies this property in Figure \ref{fig:likelihoodplot}. We compare our likelihood to the Poisson distribution that the $\hat{n}_{i,j}$ are sampled from for various values of $\lambda_i$, $n_i$, and $\hat{N}$. We also plot the likelihood used in \cite{Jethwa161207834}, which is equivalent to the mean of $P(n_i|\hat{n}_{i,j})$ over the realizations $j=1,\dots ,\hat{N}$, and we note that it does \emph{not} converge to the underlying Poisson distribution in the limit of large $\hat{N}$. We verify that this is the expected behavior by computing central moments of $P(n_i|\hat{n}_{i,j})$ for $\hat{n}_{i,j}\sim \text{Poisson}(\lambda_i)$ and using these quantities to derive central moments of $P(n_i|\hat{n}_{i,j})$ averaged over many mock observations $j=1,\dots ,\hat{N}$. For example, our analytic calculations yield $\lambda_i + 1$ and $3\lambda_i + 2$ for the mean and variance of the averaged likelihood (versus $\lambda_i$ for both quantities in the Poisson distribution), which agree with our numerical tests to high precision. Heuristically, this occurs because $P(n_i|\hat{n}_{i,j})$ is broader than (and biased with respect to) $\text{Poisson}(\lambda_i)$, so an averaged version of this likelihood cannot converge to the underlying Poisson distribution.

\section{Bayesian Evidence for Radial Scaling, Size Reduction, and Orphan Satellites}
\label{appendix}

Here we explore whether our radial scaling, satellite size reduction, and orphan models --- which were held fixed in the preceding analysis --- are favored by the classical-plus-SDSS luminosity distribution by computing Bayes factors for fits with and without these effects. In particular, we compare the Bayesian evidence for fits with $\chi=0.8$ (satellite radii scaled inward by a factor of $0.8$ relative to subhalo radii) versus $\chi=1$ (satellite radii set equal to subhalo radii), $\beta = 0$ (satellite sizes set at accretion) versus $\beta = 1$ (satellite sizes reduced based on tidal stripping), and $\mathcal{O} = 0$ (no orphan satellites) versus $\mathcal{O} = 1$ (fiducial orphan model). In all three cases, we vary $\alpha$, $\sigma_M$, $\mathcal{M}_{\rm{min}}$, and $\mathcal{B}$, and we fix all remaining parameters, except for either $\chi$, $\beta$, or $\mathcal{O}$, according to Table~\ref{tab:master}.

We calculate the evidence for each model using the bounded harmonic mean estimate, which effectively averages the inverse product of the likelihood and the prior over Markov Chain samples drawn from an ellipsoid in a high-density region of the posterior distribution. In particular, we select samples of the free parameters $\boldsymbol{\theta}$ within a fixed Mahalanobis distance \citep{Mahalanobis1936} of a point $\boldsymbol{\theta}_0$ in a high posterior density (HPD) region according to
\begin{equation}
(\boldsymbol{\theta}-\boldsymbol{\theta}_0)\boldsymbol{\Sigma}^{-1}(\boldsymbol{\theta}-\boldsymbol{\theta}_0)^T \leqslant \delta,
\label{eq:mahalanobis}
\end{equation}
where $\boldsymbol{\Sigma}$ is the covariance matrix from our MCMC fit and $\delta$ is chosen to isolate a HPD region. To estimate the (inverse of) the evidence, we average the inverse of the posterior probability for points sampled from the ellipsoid defined by Equation \ref{eq:mahalanobis}, with $\delta$ chosen to yield a particular HPD region. We then divide this quantity by the volume of the HPD region, which we compute by finding the convex hull of the sampled points. We repeat this procedure several times for values of $\delta$ chosen to yield HPD regions containing $10\%$--$25\%$ of our Markov Chain samples, which correspond to $\delta \approx 1$ for our chains; we then average the resulting values of the evidence. We have verified that our evidence estimates are converged by varying the range of HPD regions over which this mean is computed. We refer the reader to \cite{Robert09075123} and references therein for further details on this procedure.

Finally, we calculate Bayes factors $K$ by taking the ratio of the evidence for each set of fits. We find $K(\chi=0.8/\chi=1)=2.5$, $K(\beta=0/\beta=1)=0.97$, and $K(\mathcal{O}=0/\mathcal{O}=1)=0.98$; these results represent weak evidence (or, in terms of the \citealt{Jeffreys1961} scale, evidence that is ``barely worth mentioning'') against our $\chi=1$, $\beta=0$, and $\mathcal{O}=0$ models.\footnote{We have checked that these Bayes factors are monotonic functions of $\chi$, $\beta$, and $\mathcal{O}$ in order to rule out the possibility that intermediate values of these parameters (e.g., $0<\mathcal{O}<1$) are preferred.} Thus, neither radial scaling, size reduction due to tidal stripping, nor orphan satellites significantly impact the fit presented herein, and our evidence calculations justify fixing $\chi$, $\beta$, and $\mathcal{O}$ for our fit to the classical-plus-SDSS luminosity distribution. The fact that we only find weak evidence in favor of $\beta=1$ suggests that the classical-plus-SDSS luminosity distribution is fairly consistent with a subhalo--satellite size relation set at accretion. Similarly, the fact that orphans do not appreciably impact our fit hints that it is not necessary to invoke a significant amount of spurious subhalo disruption to fit the luminosity distribution of classical-plus-SDSS satellites, which correspond to subhalos with $V_{\rm{peak}}\gtrsim 20\ \rm{km\ s}^{-1}$ in our best-fit model. As discussed previously, smaller values of $\chi$ and larger values of $\mathcal{O}$ might be favored by a joint fit to the observed luminosity distribution and radial distribution of MW satellites, in which case the Bayesian evidence for models with $\chi < 1$ and $\mathcal{O}>0$ would increase accordingly.

\end{document}